\documentclass[preprint,showpacs,preprintnumbers,amsmath,amssymb,nofootinbib]{revtex4-1}

\usepackage{color}
\usepackage{graphicx}
\usepackage[dvipsnames]{xcolor}
\usepackage{hyperref}
\usepackage{subcaption}

\begin{document}

\title{One-Dimensional Simulations of the Topological Defects in a 3:1 $U(1)$ Model}


\author{Jianjun Hua$^a$}

\author{Bowen Fu$^{b,c}$}

\author{Yi-Lei Tang$^a$}
\email[]{tangylei@mail.sysu.edu.cn}

\affiliation{$^a$School of Physics, Sun Yat-Sen University, Guangzhou 510275, China}
\affiliation{$^b$Key Laboratory of Cosmology and Astrophysics (Liaoning) \& College of Sciences, Northeastern University, Shenyang 110819, China}
\affiliation{$^c$Foshan Graduate School of Innovation, Northeastern University, Foshan 528312, China}



\begin{abstract}
The domain wall is a kind of topological defect that can appear when a discrete symmetry is broken. 
If the discrete symmetry appears as an intermediate symmetry during a $U(1)$ symmetry breaking, the domain walls are connected to cosmic strings, forming walls bounded by strings.
Intuitively, the domain wall disappears if the breaking scale of the discrete symmetry is comparable to that of the $U(1)$ symmetry. 
In this paper, relying on a 3:1 $U(1)$ model, we show the detailed processes of the disappearance of the domain wall. 
Due to the existence of the non-negligible ``bias angle'' $\beta$, the relevance of the ``$Z_3$ symmetry'' and the domain wall is blurred, and thereby the evaluations of the string profiles in a hybrid wall-string network should be revised. 
We also made some preliminary calculations of the gravitational waves generated by the wall-string network created in the early universe.
\end{abstract}


\maketitle
\tableofcontents
\section{Introduction}\label{part 1}
During the phase transition in the early universe, different topological defects can form through the Kibble mechanism\cite{Kibble:1976sj}.
If the symmetry broken during the phase transition is a discrete symmetry, domain walls can form in areas between the regions broken into disconnected vacua. 
Such a structure of macroscopic scale can hardly evade the astronomical observation of our current universe\cite{Zeldovich:1974uw}, and the null result simply constrained out many of the models. 
One way to rescue such models is to embed the discrete symmetry to larger continuous symmetries, then hidden flat paths arise to connect the ``discrete'' vacua once separated by potential barriers in the field space\cite{Vilenkin:1982hm}. 
In this case, the continuous group is usually broken down step-by-step, and different types of topological defects can attach with each other, comprising the ``hybrid defects''\cite{Dunsky:2021tih}.
Two typical examples for these hybrid defects are the domain walls bounded by cosmic strings\cite{Zhou:2025zzz,Jia:2024zeu,Zhou:2026amj,Yin:2026dsr}, and the {metastable strings truncated by monopoles}\cite{Vilenkin:1984ib,
Lazarides:2024niy}.
Those hybrid defects can appear in grand unified theories\cite{Lazarides:1981fv,Kibble:1982ae,Lazarides:2023iim,Afzal:2023kqs,Maji:2024pll,Maji:2025thf,Maji:2026nkz}, two Higgs doublet models\cite{Fu:2024rsm} and axion-like models\cite{Stecker:1982ws,Eto:2023aqr,Bao:2024bws,Lee:2024toz,Dvali:2025ivw}. 
The phenomenologies are also discussed in Ref.~\cite{Dunsky:2021tih,Tranchedone:2026lav}.

Unlike the pure discrete symmetry scenario, the domain wall attaching the cosmic string is fundamentally unstable. Quantum tunnelling can tear the domain wall plane, and eventually, all the domain walls will disappear after a sufficiently long time elapses. 
The key problem lies in the time scale, depending mainly on the vacuum expectation values (VEVs) of the charged scalar fields, as well as their ratio. Usually, a larger VEV, nominated as $v_1$, connecting with the breaking scale of the continuous symmetry arises, followed by another smaller $v_2$ breaking the remaining discrete symmetry. In the previous researches, $|v_1| \gg |v_2|$ is usually appointed, and the domain wall solution is acquired in the effective theory by integrating out the Higgs degrees of freedom corresponding to $v_1$. Intuitively speaking, relaxing such a condition destroys the domain wall solution, but how does this happen? If we decrease the $v_1$ to the scale $|v_1| \sim |v_2|$, or even in a reversed $|v_1| < |v_2|$, will the domain wall still exist? If so, is the domain wall at least metastable?

Unlike the usual $U(1)\rightarrow Z_2$ model described in the literature\cite{Wu:2022tpe,Fu:2024jhu,Li:2025gld,Wu:2022stu,Barreto:2026igt,Jueid:2023cgp,Chen:2026fod}, in this paper, we alternatively rely on a gauged $3:1$ $U(1)$ model as an example, which is originally designed for a $U(1)\rightarrow Z_3$ breaking process. Similarly, examples for the domain-wall models beyond the $Z_2$ symmetry can be seen in Ref.~\cite{Jobu:2025tto,Fu:2025qhf}. In this paper, we are particularly interested in the $|v_1| \sim |v_2|$ situation, so we lower the ``continuous symmetry breaking'' scale by decreasing the corresponding $v_1$ to scrutinize the domain wall solution status. 
Comparing the results from both the ``shooting method'' and the ``path deformation algorithm''\cite{Wainwright:2011kj,Chen:2020wvu,Chen:2020soj}, we trace the phases of the domain walls as $v_1$ varies, and the disappearance processes of the domain wall solution are presented. Besides, the two vacua separated by the domain wall are no longer connected by the rigorous $Z_3$ transformation. We introduce the ``bias angle'' $\beta$ to describe such a deviation. In the $v_1 \rightarrow \infty$ limit, the bias angle vanishes, and the usual $U(1)\rightarrow Z_3$ scenario is restored naturally. In the early universe, if described by such a model, the hybrid wall-string network arises. Preliminary calculations of the corresponding gravitational wave signals are calculated, offering some opportunities to test the scenarios.

In Sec. \ref{part 2}, the Lagrangian is introduced, and the equations of motion(EOMs) are derived and presented. In Sec. \ref{part 3}, the topological defects incorporated in the model are described briefly, the defect networks    and their evolution are included as well. The calculation algorithms, techniques and numerical results of defects are addressed in detail in Sec. \ref{Algorithm_Techniques_Calc}. Sec. \ref{part 5} introduces the corresponding gravitational wave spectrum formulas of defect annihilation and give the gravitational spectrum. Finally, in Sec. \ref{part 6} we summarise the paper.

\section{Model Description and the Basic Equations of Motion}\label{part 2}

In this paper, we consider a model with two complex scalars $\phi_{1,2}$ charged under the same $U(1)$ gauge group.
The ratio of the scalar charges are $Q_1 : Q_2=3:1$. The corresponding gauge field is denoted by $A_{\mu}$. The Lagrangian is given by
\begin{eqnarray}\label{model}
\mathcal{L}=-\frac{1}{4} F_{\mu \nu} F^{\mu \nu} + \left | D_{\mu} \phi_1 \right |^2 +\left | D_{\mu} \phi_2 \right |^2 -V\left(\phi_1 ,\phi_2\right),
\end{eqnarray}
where $F_{\mu \nu} = \partial_{\mu} A_{\nu} - \partial_{\nu} A_{\mu}$, $D_{\mu}  = \partial_{\mu} - i g\, Q A_{\mu}$ with $g$ the $U(1)$ gauge coupling and $Q$ the $U(1)$ charge.
The most general renormalizable scalar potential is 
\begin{eqnarray}\label{potential}
V\left(\phi_1 ,\phi_2\right) = -\mu_1^2 \left| \phi_1 \right|^2-\mu_2^2 \left|  \phi_2 \right|^2 + \lambda_1\left| \phi_1 \right|^4 + \lambda_2\left| \phi_2 \right|^4 + \lambda_{12}\left| \phi_1 \right|^2\left| \phi_2 \right|^2-( \lambda_{3} \phi_1^*\phi_2^3+ \text{h.c.}).
\end{eqnarray}
Here, $\mu_{1,2}^2$ are the mass parameters, and $\lambda_{1,2,12,3}$ are the corresponding coupling constants. Note that the Hermitianity of the action guarantees the $\mu_{1,2}$ and $\lambda_{1,2,12}$ to be real, while $\lambda_3$ can be complex. 
For simplicity, we further impose the $CP$-conservation so that $\lambda_3$ is real and set the gauge coupling $g=1$ in the rest part of the paper. Also notice that flipping the signs of either $\phi_{1,2}$ reverses the $\lambda_3$, while other couplings and mass terms remain intact. Therefore, without loss of generality we appoint $\lambda_3 > 0$.

The complex scalar fields can be decomposed into their real and imaginary parts
\begin{eqnarray}
    \phi_{a} = \frac{\phi_{a r} + i \phi_{a i}}{\sqrt{2}},  \label{phi12_Expansion}
\end{eqnarray}
where $\phi_{ar}$, $\phi_{ai}$ are the real scalar components.
With this basis, the scalar potential reads
\begin{eqnarray}
    V &=& - \frac{\mu_1^2}{2}\left(\phi_{1i}^2 + \phi_{1r}^2\right) - \frac{\mu_2^2}{2}\left(\phi_{2i}^2+\phi_{2r}^2\right) + \frac{\lambda_1}{4}\left(\phi_{1i}^2 + \phi_{1r}^2\right)^2 + \frac{\lambda_2}{4}\left(\phi_{2i}^2 + \phi_{2r}^2\right)^2 \nonumber \\
    && + \frac{\lambda_{12}}{4}\left(\phi_{1i}^2+\phi_{1r}^2\right)\left(\phi_{2i}^2+\phi_{2r}^2\right) - \frac{\lambda_3}{2}\left(\phi_{1r}\phi_{2r}^3 - \phi_{1i}\phi_{2i}^3 + 3\phi_{2r}^2\phi_{2i}(\phi_{1i}\phi_{2r} - \phi_{1r}\phi_{2i}) \right)\,.
\end{eqnarray}
The minimization condition of the scalar potential is given by
\begin{eqnarray}
    \frac{\partial V}{\partial \phi_{1r}} = \frac{\partial V}{\partial \phi_{1i}} = \frac{\partial V}{\partial \phi_{2r}} = \frac{\partial V}{\partial \phi_{2i}} = 0\,. \label{StationaryEquations}
\end{eqnarray}
The alternative expressions for the complex scalar fields are with the polar basis, denoted by  
\begin{eqnarray}
    \phi_{a} = \frac{\Phi_a}{\sqrt{2}} e^{i\varphi_a}, 
\end{eqnarray}
where $\Phi_\alpha\geq 0$. With these the scalar potential can be written as 
\begin{eqnarray}
    V=-\frac{\mu_1^2}{2} \Phi_1^2 - \frac{\mu_2^2}{2} \Phi_2^2 + \frac{\lambda_1}{4}\Phi_1^4 + \frac{\lambda_2}{4} \Phi_2^4 + \frac{\lambda_{12}}{4}\Phi_1^2\Phi_2^2 - \frac{1}{2}\lambda_3 \Phi_1 \Phi_2^3 \cos(\varphi_1-3\varphi_2)
\end{eqnarray}

After the symmetry breaks completely, the vacuum expectation values (VEVs) of the scalars can be written as $\langle \phi_i \rangle = \frac{v_i}{\sqrt{2}} e^{i \varphi_i}$, where $v_i$ are positive real numbers and $\varphi_i$ indicates the corresponding the phases. 
Since $\lambda_3>0$, it is clear that the potential minima should satisfy the condition $\varphi_1=3\varphi_2$.

\begin{figure}[t!]
    \centering
    \includegraphics[width=0.7\linewidth]{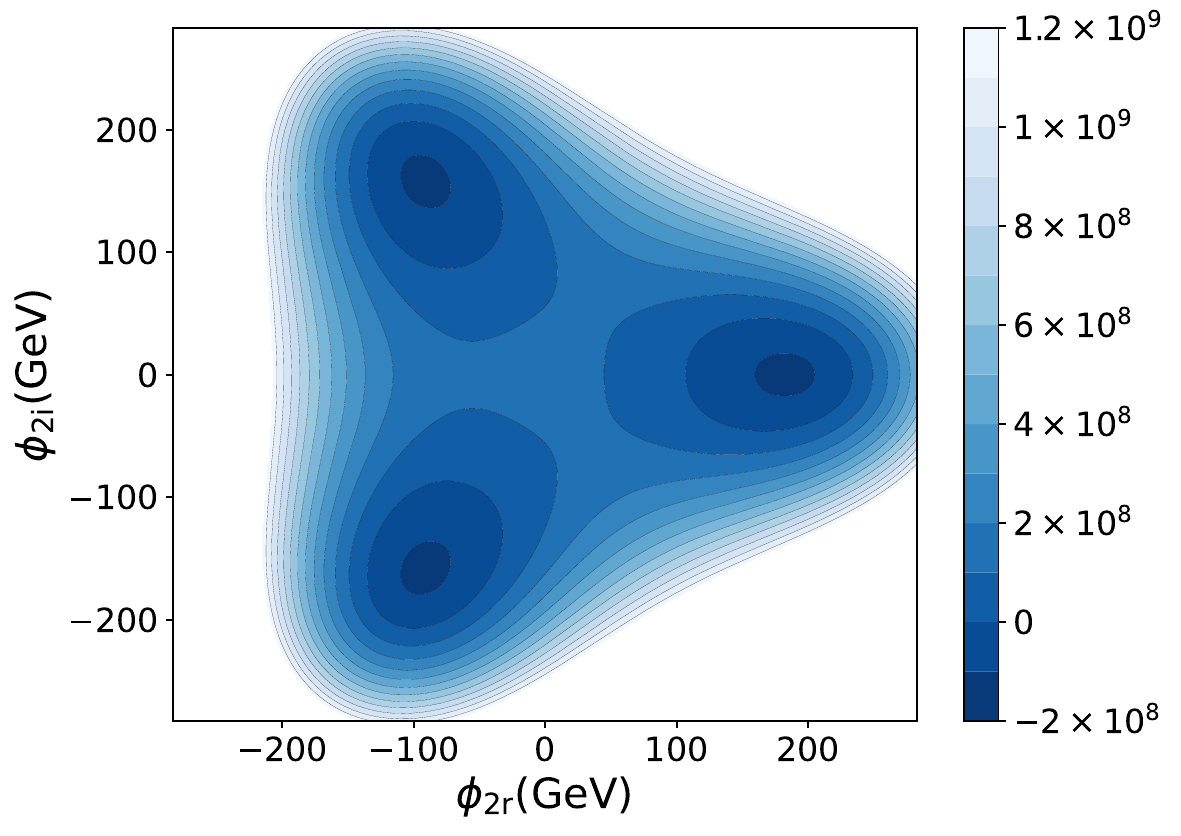}
    \caption{\label{phi2_potential} The projected potential $V\left(\frac{v_1}{\sqrt{2}}, \phi_2\right)$ in the $\phi_{2r}-\phi_{2i}$ plane with the first parameter fixed at $v_1$. The adopted parameters are $v_1=100\sqrt{2}\text{GeV}$, $v_2=130\sqrt{2}\text{GeV}$, $\lambda_1=1$, $\lambda_2=1.6$, $\lambda_{12}=0$, $\lambda_3=1$. }
\end{figure}
Once the VEVs $v_1$ and $v_2$ are determined, there arises a series of degenerate vacua linked by $U(1)$ transformation. In the case of the hierarchical VEVs $v_1 \gg v_2$, the whole symmetry breaking processes of the $U(1)$ group can be separated into two sub-processes: $U(1) \rightarrow Z_3$ and $Z_3 \rightarrow \lbrace e \rbrace$, where $\lbrace e \rbrace$ refers to the trivial group. 
When the $\phi_1$ obtains a real VEV $\langle\phi_1\rangle=\frac{v_1}{\sqrt{2}}$, the potential of the $\phi_2$ is shown in Fig.\ref{phi2_potential}, in which three degenerate minima line in the three vertices of an equilateral triangle at $\langle \phi_2 \rangle = \frac{v_2}{\sqrt{2}} e^{i \frac{2 n \pi}{3}}$, with $n=0,1,2$ connected by the $U(1)$ transformations in the full four-dimensional field space, but disconnected within the two-dimensional section confined by $\langle \phi_1 \rangle =\frac{v_1}{\sqrt{2}}$.
If different parts of the universe fall into different $\langle \phi_2 \rangle$ before $\langle \phi_1 \rangle$ acquires a universal VEV $\langle \phi_1 \rangle=\frac{v_1}{\sqrt{2}}$, domain walls arise as the boundaries among these different vacua (See more details in Sec.~\ref{part 3}). In the case of $v_1 \gg v_2$, $\phi_1$ can be integrated out after the $U(1)$ breaks, leaving a $Z_3$-symmetric theory of the $\phi_2$ with which the domain walls can arise.
However, in this paper, we are particularly interested in a different scenario in which $v_1$ and $v_2$ are of a similar scale, where all degrees of freedom become indispensable.

\section{Topological defects}\label{part 3}

Topological defects, including monopoles, cosmic strings, and domain walls, can form through the Kibble mechanism during the early universe phase transitions \cite{Kibble:1976sj,Kibble:1982ae,Kibble:1982dd}. 
In the model considered here, cosmic strings can arise as the $U(1)$ symmetry breaks, which are cylindrical solutions to the classical field EOMs. 
Along the tangential direction of the strings, the fields vary along the flat direction of the potential in the four-dimensional field space. 
The main contribution to the string tension is from the gradient energy of the fields. 

Apart from the string solution, it is possible to form a domain wall between the two vacua. 
When $\langle \phi_1 \rangle \gg \langle \phi_2 \rangle$, domain walls are formed between the vacua linked by one of the $Z_3$ subgroup in the $U(1)$. For an example, domain walls can arise between the vacua $\langle \phi_2 \rangle = \frac{v_2}{\sqrt{2}}$ and $\langle \phi_2 \rangle = \frac{v_2}{\sqrt{2}} e^{i \frac{2 \pi}{3}}$, sharing a common $\langle \phi_1 \rangle = \frac{v_1}{\sqrt{2}}$ remaining nearly fixed inside the wall. In this case, the tensions of the string and the wall are incomparable, however, when the wall and string tensions become compatible when $\langle \phi_1 \rangle \sim \langle \phi_2 \rangle$, possibilities arise for the formation of a mixed structure of walls and strings. In such a case, it becomes less precise to claim that the two vacua on both sides of the domain wall respect a discrete subgroup. Specifically, relying on the model introduced in this paper, we will see that if the $v_1$ gradually decreases, the disturbance of the $\phi_1$ in the vicinity of the domain wall becomes non-negligible, and finally the phases $\varphi_1$ of the two vacua separated by the domain wall gradually deviate. Finally, $\varphi_1$ is no longer fixed as in the $v_1 \gg v_2$ limit, and additional difference of the phases from the $Z_3$ values shows up on the two sides of the wall. Such a phase difference can become so significant that the $Z_3$ symmetry connecting both the vacua along the domain wall is lost.

In order to calculate the profile of the topological defects, we have to write down the EOMs derived from the Euler-Lagrange equation,
\begin{eqnarray}\label{general EOM}
& & D^{2} \phi_{1,2}+ \frac{\partial V}{\partial \phi^{*}_{1,2}}=0 \nonumber \\
& & D^{*2} \phi_{1,2}^{*}+ \frac{\partial V}{\partial \phi_{1,2}}=0 \nonumber \\
& & \partial_{\nu} F^{\mu \nu}+\sum _{k=1,2} i Q_k \left[\phi_{k}^{*} D^{\mu} \phi_{k}-\left(D^{\mu} \phi_{k} \right)^{*}\phi_{k}\right]=0, 
\end{eqnarray}
where $D^{2}=D_{\mu} D^{\mu}$.
In this paper, we focus on the static one-dimensional solutions to Eq.~\eqref{general EOM} based upon the symmetry ansatz. 
The detailed one-dimensional differential equations, as well as the boundary conditions of both domain walls and cosmic strings, are described in the following two subsections, followed by the qualitative analysis of the hybrid defects by simply combining the two structures together in the last subsection.

\subsection{Domain Wall Solution}\label{domain wall part}

The static domain wall solution is macroscopic in two space directions and microscopic in one space direction. 
Thus, the EOMs can be simplified along the microscopic direction, chosen as the $z$-axis without loss of generality in this paper.

On the other hand, due to the microscopic homogeneity of the domain wall, each vector along the $x$-$y$ plane is equivalent and thus $A_{1,2}$ vanish. 
The non-zero $A_3$ can also be removed easily by a simple gauge transformation $A_3 \rightarrow A_3 + \partial_z \phi_A$, where $\phi_A = -\int_{z_0}^z A_3(0, 0, z^{\prime}) dz^{\prime}$. 
$A_0$, being the electric potential, also vanishes due to the time independence of the $\phi_{1,2}$ fields. As a result, the general static EOMs Eq.~\eqref{general EOM} can be simplified into 
\begin{eqnarray}\label{DW equation}
\phi_{1,2}^{\prime \prime}=\frac{\partial V(\phi_1,\phi_2)}{\partial \phi^*_{1,2}} ,
\end{eqnarray}
where $X^\prime$ refers to ${dX}/{dz}$ and the conjugate equations are omitted. 
Substituting the $V(\phi_1, \phi_2)$ by Eq.~\eqref{potential}, it turns out 
\begin{eqnarray}
& & \phi^{\prime \prime}_1=-\mu_1^2 \phi_1+2\lambda_1 \left|\phi_1\right|^2 \phi_1+\lambda_{12} \left|\phi_2\right|^2\phi_1-\lambda_{3}\phi_2^{*3}\nonumber \\
& & \phi^{\prime \prime}_2=-\mu_2^2 \phi_2+2\lambda_2 \left|\phi_2\right|^2 \phi_2+\lambda_{12} \left|\phi_1\right|^2\phi_2-3 \lambda_{3} \phi_1^{*}\phi_2^{2}. \label{DomainWallEOM}
\end{eqnarray}
Without the loss of generality, we choose the vacuum on one side of the wall as 
\begin{eqnarray}
    \lim_{z \rightarrow -\infty} \phi_1 = \frac{v_1}{\sqrt{2}},
        \lim_{z \rightarrow -\infty} \phi_2 = \frac{v_2}{\sqrt{2}} e^{i \frac{2 \pi}{3}},
\end{eqnarray}
while the vacuum on the other side is chosen as
\begin{equation}
        \lim_{z \rightarrow +\infty} \phi_1 = \frac{v_1}{\sqrt{2}} e^{i \beta}, 
        \lim\limits_{z \rightarrow +\infty} \phi_2 = \frac{v_2}{\sqrt{2}} e^{i \frac{\beta}{3}}. \label{DomainWallBoundaryCondition}
\end{equation}
The phase $\beta$ indicates the deviation of the two asymptotic vacua from a rigorous $Z_3$ transformation, which is referred to as the ``bias angle'' in this paper.
$\beta$ is a free parameter that can be determined by minimising the tension of the domain wall, which is addressed by
\begin{eqnarray}
\sigma=\int_{-\infty}^{+\infty}\left [ \sum_{i=1,2}\left( \frac{\partial \phi_{i}}{\partial z} \right)^2+V(\phi_1,\phi_2)\right]dz. \label{Tensor_Domainwall}
\end{eqnarray}
Particularly, when $\beta = 2 \pi$, there always exists a trivial solution, indicating the homogeneous vacuum with no wall, which should be avoided during the pragmatic calculation processes.

\subsection{String Solution} \label{StringSolutionSubsec}

The cosmic string is a one-dimensional macroscopic object. 
From the microscopic perspective, it admits a two-dimensional rotational symmetry, which can
in general be written in the form of
\begin{eqnarray}
\phi_{i}= f_{i}(r) e^{i n_{i} \theta_{i}},\quad A=A_{\theta}\vec{e}_{\theta},\quad A_{\theta}=\frac{1}{r} a(r) , \label{String_Field_Expansion}
\end{eqnarray}
in the plane perpendicular to the string with a polar coordinate, where $n_i$ are the ``winding numbers'' of the strings, $\vec{e}_{\theta}$ indicates the unit vector along the tangential direction. Along a circle path centred at the origin, the field values are connected through the gauge transformations, thus $n_1:n_2 = Q_1:Q_2 = 3:1$.
Then the static EOMs become
\begin{eqnarray}
    \left( \frac{d^2}{d r^2} + \frac{1}{r} \frac{d}{dr} - \frac{n_1^2}{r^2} - Q_1^2 A_{\theta}^2 + \frac{2 n_1 Q_1 A_{\theta}}{r} \right) f_1 &=& -\mu_1^2 f_1 + 2\lambda_1 f_1^2 f_1+\lambda_{12} f_2^2 f_1-\lambda_{3} f_2^{3}  \nonumber \\
    \left( \frac{d^2}{d r^2} + \frac{1}{r} \frac{d}{dr} - \frac{n_2^2}{r^2} - Q_2^2 A_{\theta}^2 + \frac{2 n_2 Q_2 A_{\theta}}{r} \right) f_2 &=& -\mu_2^2 f_2+2\lambda_2 f_2^2 f_2+\lambda_{12} f_1^2 f_2-3 \lambda_{3} f_1 f_2^{2} \nonumber \\
    \left(\partial_{r}^{2}+\frac{1}{r}\partial_{r}-\frac{1}{r^2}\right){a} &=& 2 \sum_{i=1,2} \left[Q_{i} \left|\phi_{i}\right|^2\left(n_{i}-Q_{i} a\right)- i{r} Q_{i}\phi_{i}^{*}{\frac{d\phi_i}{dr}}\right]  . \label{CosmicStringEOM}
\end{eqnarray}
The boundary conditions are given by
\begin{eqnarray}\label{string boundary condition}
&f_{i}(r)\Big|_{r=0} = 0\,,\quad 
&a(r)\Big|_{r=0} = 0\,,\\
& f_{i}(r)\Big|_{r\rightarrow\infty} = \frac{v_{i}}{\sqrt{2}}\,,\quad 
&a(r)\Big|_{r\rightarrow\infty} = {\frac{n_i}{Q_i}}. 
\end{eqnarray}

\begin{figure}[t!]
\centering
    \begin{subfigure}[c]{0.3\textwidth}   
        \centering
        \includegraphics[width=\linewidth]{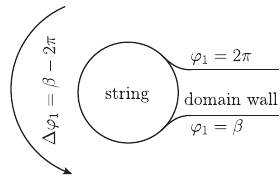}
        \caption{}
    \end{subfigure}
    \hspace{0.6cm}  
    \begin{subfigure}[c]{0.3\textwidth}   
        \centering
        \includegraphics[width=\linewidth]{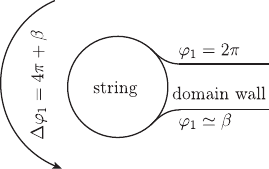}
        \caption{}
    \end{subfigure}

\caption{\label{sketch_rim} The sketches of the two simplest ``rims", with which the wall ends. The difference lies at the $n$ values in (\ref{Fractional_n1}), for which $n=0$, $1$ are adopted, corresponding to the left and right panels respectively. In both panels phases $\varphi_2=\varphi_1/3$, which are neglected in the sketches.}
\end{figure}

The periodic boundary conditions of a complete string require that both $n_1$ and $n_2$ are integers. 
Therefore, the ground state of a cosmic string requires $n_1=3$ and $n_2=1$ for both $v_1 \neq 0$ and $v_2 \neq 0$. 
However, in the $v_1 \gg v_2$ case, indicating a chain of stepwise breaking processes, $n_1 = 1$ can be satisfied during the period when $\langle \phi_2 \rangle =0$, before $\langle \phi_2 \rangle$ acquires a nonzero VEV for the domain walls thereby arising between the strings. 
The domain walls might also collapse through the stretch of a string-bounded hole nucleating on it by the tunnelling effects. 
Both these hybrid defect structures had been summarised in the Ref.~\cite{Dunsky:2021tih}, and in this paper, we further point out that around the string, $\phi_2$ changes abruptly along both sides of the domain wall, forming a discontinuity so that $n_2$ is no longer an integer. 
Furthermore, due to the existence of the ``bias angle'' $\beta$ introduced in (\ref{DomainWallBoundaryCondition}), when going through the domain wall, the phase of $\phi_1$ no longer continues. 
Therefore, in order to connect the phases of the $\phi_i$ on both sides of the domain wall, suppose the wall lies at $\theta=0$, then $\phi_2(\theta = 2 \pi) = e^{i \frac{n_1}{3} (2 \pi)} = e^{i \frac{2 \pi - \beta}{3}} = \phi_2(\theta = 0) e^{i \frac{2 \pi - \beta}{3} }$ requires 
\begin{eqnarray}
    n_1 = 1+ 3n - \frac{\beta}{2 \pi},~~n \in \mathbb{Z}, \label{Fractional_n1}
\end{eqnarray}
which also becomes fractional. Interestingly, such fractional winding numbers also arise in condensed matter physics area, with an example to be found in Ref.~\cite{Makinen:2018ltj}. Two examples of the profiles of such kinds of hybrid structures are sketched in Fig.~\ref{sketch_rim}.

The tension of the string, or the linear energy density, can be evaluated by the following formula\cite{Hindmarsh:1994re,Dunsky:2021tih},
\begin{eqnarray}
& & \mu=\int rdrd \theta \sum_{i=1,2}\left[ \left|\frac{\partial \phi_{i}}{\partial r} \right|^2+\left|\frac{1}{r}\frac{\partial\phi_{i}}{\partial \theta}-i Q_{i} A_{\theta} \phi_{i}\right|^2+ V\left(\phi_{1},\phi_{2}\right)+\frac{B^2}{2}\right], \label{Tensor_CosmicString}
\end{eqnarray}
where $B$ is magnetic field related to $A^{\mu}$.

\subsection{merge of walls and strings}\label{hybrid defects}

\begin{figure}[htbp]
    \centering
    \begin{subfigure}[c]{0.4\textwidth}   
        \centering
        \includegraphics[width=\linewidth]{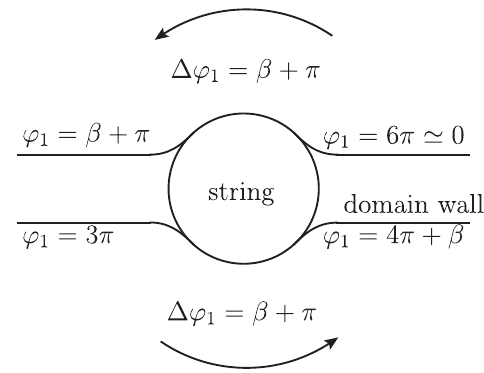}\hspace{0.4cm}
        \caption{}
    \end{subfigure}
    \hspace{0.3cm}  
    \begin{subfigure}[c]{0.3\textwidth}
        \centering
        \includegraphics[width=\linewidth]{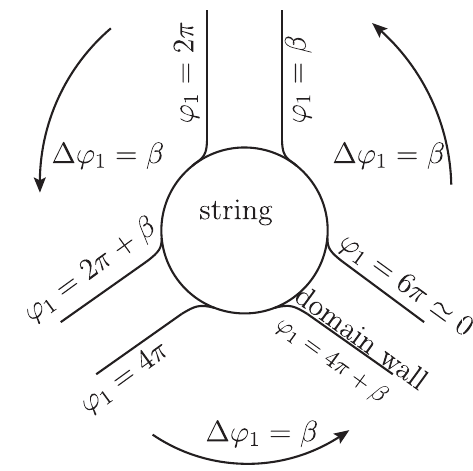}
        \caption{} \label{Three_Conjunctions_Sketch}
    \end{subfigure}
    \caption{ The simplest (a)``Two-conjunction''  and  (b) ``three-conjunction''. } \label{sketch_Conjunctions}
\end{figure}

In our model, as in the 2:1 U(1) model in the Ref.~\cite{Kibble:1982dd,Preskill:1992ck,Dunsky:2021tih}, there are hybrid string-wall structures, which are called the ``rims'' of domain walls. 
Two examples are sketched in Fig.~\ref{sketch_rim}. 
Besides, ``conjunctions'' might also arise. 
Typical examples are sketched in Fig.~\ref{sketch_Conjunctions}, where more than two domain walls are connected to converge into a strip-shaped object. 
All these structures can co-exist, forming a complicated network. 
In this paper, we are particularly interested in the $v_1 \sim v_2$ parameter space, so the existence of the domain wall disturbs the rotational symmetry around the strip-shaped object at the end of each wall-sheet.
Precise solutions to these hybrid structures require at least two-dimensional simulations, which are beyond the scope of this paper. 
A simple alternative estimation is to regard these strips as strings. 
Again, as in the ``rim'' case, for the ``conjunctions'' the existence of the bias angle $\beta$ in the domain walls makes the phase increments allocated around the ``string'' no longer integers times $2 \pi$, making both the effective $n_{1,2}$ fractional, as has been illustrated in (\ref{Fractional_n1}).

\begin{figure}[t!]
\includegraphics[width=0.6\textwidth]{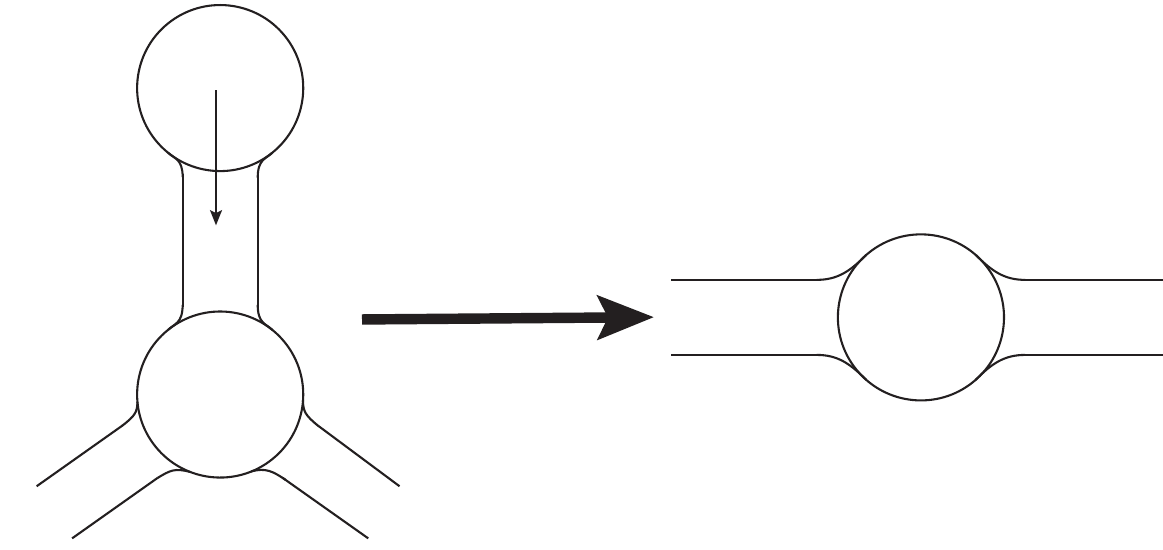}
\caption{\label{merge_3DW} Three-conjunction merges with a rim pulled by the tension of the wall, forming a two-conjunction.}
\end{figure}

\begin{figure}[t!]
\includegraphics[width=0.8\textwidth]{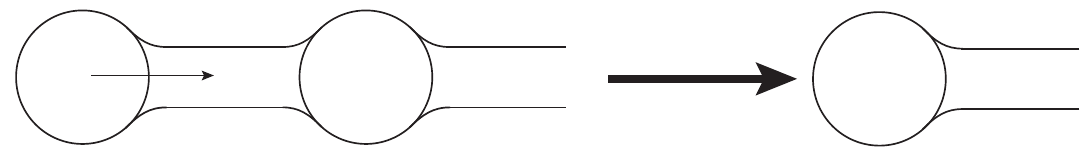}
\caption{\label{merge_2DW} Two-conjunction merges with a rim, forming another rim. }
\end{figure}

\begin{figure}[t!]
\centering

\includegraphics[width=0.8\textwidth]{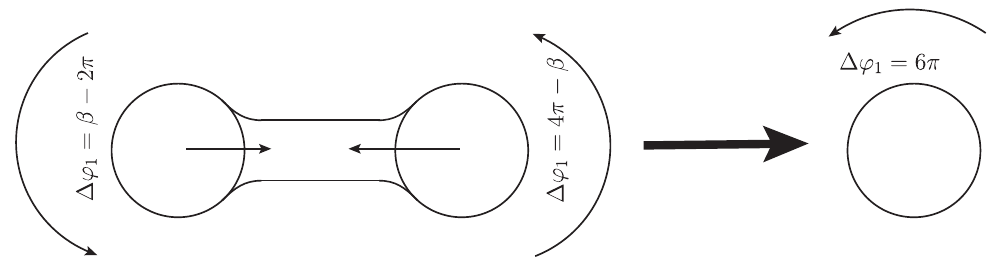}
\vspace{1em}

\includegraphics[width=0.8\textwidth]{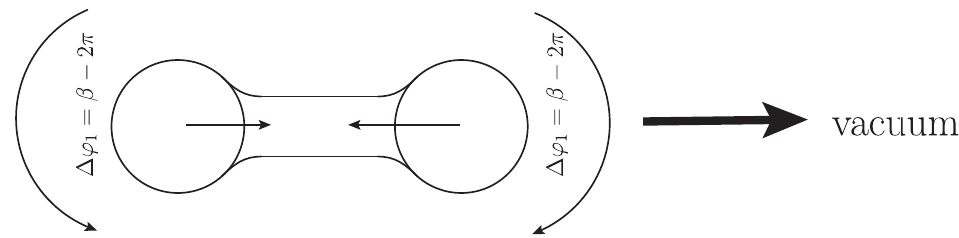} 
\caption{\label{merge_Rims} Two rims merge to a string or annihilate. The results depend on the participated rim types, thereby the total phase increment $\Delta \varphi_1$ surrounding the overall structure.}
\end{figure}

Once the string-wall network is generated, the strings are pulled due to the tension of the walls. 
Space symmetries of the conjunctions in Fig.~\ref{sketch_Conjunctions} balance the forces of the string, while the rims in Fig.~\ref{sketch_rim} are dragged. 
Besides the simple shrinking of a rim loop as described in Ref.~\cite{Dunsky:2021tih}, rims might collide with conjunctions bounding the same wall-sheet to fuse into another type of ``conjunction'' or ``rim''. 
Some examples are sketched in Fig.~\ref{merge_3DW} and \ref{merge_2DW}. 
Finally, two rims can be pulled together by the domain wall connecting them, either fuse into a pure cosmic string, or disappear into the vacuum, depending on the effective $n_1$ values of the two rims, as sketched in Fig.~\ref{merge_Rims}.

\section{Algorithms, Techniques and Results for the Topological Defects}\label{Algorithm_Techniques_Calc}

In order to acquire the profiles of the domain walls and the cosmic strings, we have to solve (\ref{DomainWallEOM}) and (\ref{CosmicStringEOM}) under the boundary conditions (\ref{DomainWallBoundaryCondition}) and (\ref{string boundary condition}), respectively. We apply both the classical shooting-method and the path deformation algorithm  \cite{Wainwright:2011kj,Chen:2020wvu,Chen:2020soj} to find the ``eigen functions'' satisfying the boundary conditions. For the shooting-method we make our own programs relying on the Python-based differential equation solvers. For the path deformation algorithm, we adopt the modified Cosmotransitions \cite{Wainwright:2011kj,Chen:2020wvu,Chen:2020soj} code. The shooting method tends to be more precise in the existence of the bias angle, while the path deformation algorithm runs faster, and can help us verify the stability of the solutions. We are going to compare both the results.

Practically, it is difficult for the numerical evaluations to touch the place of the infinite distance, and the shooting-method also suffer from the possible divergences induced from the rounding errors in the remote space. Therefore, when applying the shooting-method, we have to solve the equations analytically beyond a sufficiently far distance before ``matching'' the numerical results with these asymptotic analytical solutions at some point.  Fortunately, in a remote area, all the non-linear terms in the EOMs become negligible, making it easier for us to solve these approximated linear equations. For the path deformation algorithm to solve the domain wall, ``matching'' processes are exempted, however the bias angle $\beta$ should be appointed as a parameter from the beginning, which is unknown before the solution is acquired. Hence, we have to search through all possible $0 < \beta < 2 \pi \times 3$ for the extrema of the (\ref{Tensor_Domainwall}) to find the true values of $\beta$, upsurging the time consumption to a scale comparable with the shooting-method. In this paper, we adopt both these methods for a comparison. In the following of this section, we mainly illustrate the details about the numerical processes, including the asymptotic solutions at the infinite distance for the shooting curve to match, and the ``target functions'' to measure the convergence for the shooting method. Discussions about the parameter space will be presented, followed by parts of the numerical results.

\subsection{Asymptotic Solutions}

\subsubsection{Asymptotic Solutions for the Domain Wall} \label{As_Wall}

As $z \rightarrow \pm \infty$, expanded around the potential minimum according to (\ref{phi12_Expansion}) where $\varphi = \frac{2 \pi}{3}$ or $\varphi = \beta$ are given by (\ref{DomainWallBoundaryCondition}), the constant terms in (\ref{DomainWallEOM}) vanish automatically which is guaranteed by the stationary equations (\ref{StationaryEquations}). Preserving only the linear terms as an approximation, we can change the form of (\ref{DomainWallEOM}) into
\begin{eqnarray}
    \Phi^{\prime \prime} = M^2_{4 \times 4} \Phi, \label{EOM_Linear_Part_DomainWall}
\end{eqnarray}
where $\Phi = [\phi_{1r}, \phi_{1i}, \phi_{2r}, \phi_{2i}]^T$ is a column vector
, and $M^2_{4 \times 4}$ is the $4 \times 4$ squared mass matrix relative to the $(v_1, v_2 e^{i \frac{2 \pi}{3}})$ or $(v_1 e^{i \beta}, v_2 e^{i \frac{\beta}{3}})$ vacua. Then, $M^2_{4 \times 4}$ is diagonalized by an orthogonal matrix $V$ satisfying $V V^T = V^T V = I$,
\begin{eqnarray}
    V M^2_{4 \times 4} V^T = \text{diag}[m_1^2, m_2^2, m_3^2, 0], \label{Diag_M44}
\end{eqnarray}
where the stability of the vacuum requires that $m_{1,2,3,4}^2>0$, and the zero eigenvalue indicates the Goldstone degree of freedom, which is the ``flat'' direction of the effective potential. Define
\begin{eqnarray}
    \tilde{\Phi} = V^{T} \Phi = [\tilde{\phi}_1, \tilde{\phi}_2, \tilde{\phi}_3, G] \label{shift to NGB direction},
\end{eqnarray}
so (\ref{EOM_Linear_Part_DomainWall}) becomes
\begin{eqnarray}
\left\lbrace
\begin{array}{l}
    \tilde{\phi}_i^{\prime \prime} = m_i^2 \tilde{\phi}_i, (i=1,2,3)\\
    G^{\prime \prime} = 0
\end{array} \right. .
\end{eqnarray}
Solving these equations, we have 
\begin{eqnarray}
    \tilde{\phi}_i &=& C_{i}^+ e^{m_i z} + C_{i}^- e^{-m_i(z-z_0)}, \nonumber \\
    G &=& G_0 + { G_1} (z-z_0), \label{Asymp_Domain_Wall}
\end{eqnarray}
where $C_{i}^{\pm}$ and $G_0$, ${G_1}$ are arbitrary constants, and $z_0>0$ is the matching point between the numerical algorithm and the asymptotic solution in the $z \rightarrow \infty$ limit, which should be sufficiently large. At the ``initial point'' where we start to shoot in $z = - z_0$, the $\lim\limits_{z \rightarrow -\infty} \tilde{\Phi} = 0$ condition requires $C_{i}^- = 0$ and $G_{0,1} = 0$, so $C_i^+$, as well as the bias angle $\beta$ becomes the unknown parameters for the numerical root finder procedures to solve. $-z_0 < z < z_0$ is the range in which the numerical solution processes are performed, until $z=z_0$, we then need a smooth transition to the asymptotic solutions, alternatively requiring $C_{i}^+ = 0$ and ${G_1}=0$ at this point, becoming the zero conditions input into the root finders to solve the reasonable initial $C_{1,2}^+$ and $\beta$. $G_0$ also has to vanish if we expand the fields relative to the vacuum with the correct bias angle $\beta$, however practically the precise value of $\beta$ is difficult for us to acquire before hand, therefore we tolerate a small nonzero $G_0$ if $G_0 \ll v_{1,2}$, and finally absorb its contribution to correct the $\beta$ by 
\begin{eqnarray}
    \delta \beta \approx \frac{3 G_0(z_0)}{\sqrt{9 v_1^2+v_2^2}}, \label{delta_beta_correction}
\end{eqnarray}
in which $G_0$ finally vanishes. Practically, in most cases, we find out that $G_0 \lesssim v_{1,2} \times 10^{-3}$, verifying the validity of this trick.

\subsubsection{Asymptotic Solutions for the Cosmic String} \label{Asymp_String_Discussions}

Starting from $r=0$ when solving (\ref{CosmicStringEOM}) encounters a $\frac{1}{r}$ singularity in the equations. Alternatively one has to begin at a small $r_{s}>0$, and the solutions of $f_{1,2}$ there are estimated to be $f_i(r) \approx C_i r^{\alpha_i}$ where $C_{1,2}$ are constants. Substitute these into (\ref{CosmicStringEOM}), and neglect the higher order terms, we have
\begin{eqnarray}
    C_{i} \left[\alpha_i \left(\alpha_i-1\right)+\alpha_i-n_{i}^{2}\right] r^{\alpha_i-2}=0, \label{f_Equation}
\end{eqnarray}
This implies $\alpha_i=\left| n_i \right|$. The gauge part of the (\ref{CosmicStringEOM}) therefore becomes
\begin{eqnarray}
    \left(\partial_{r}^{2}+\frac{1}{r}\partial_{r}-\frac{1}{r^2}\right)A^{\theta} &=& 2 Q_{2} |C_2|^2 r^{2|n_{2}|}\frac{n_{2}}{r}, \label{r0_asymp_A}
\end{eqnarray}
at the $r \rightarrow 0$ limit. Here the $\propto r^{2 \alpha_i} A^{\theta}$ and $\propto r^{2 \alpha_1} \frac{n_1}{r}$ terms are neglected due to their higher order in the expansion. Solving (\ref{f_Equation}) and (\ref{r0_asymp_A}) gives
\begin{eqnarray}
    f_i(r)=C_i r^{|n_i|} ,\quad a(r)=C_0 r^2 + \frac{2 |C_2|^2 Q_2 |n_{2}|}{4|n_{2}|^2+4|n_{2}|} r^{2|n_{2}|+2}, \label{r0_Asymptot}
\end{eqnarray}
where the definition of $a(r)$ can be found in (\ref{String_Field_Expansion}), and $C_0$ is another constant. Here $C_{0,1,2}$ are the unknown parameters for the equation solving procedures to find out.

For the $r\rightarrow \infty$ side, it has been known that $|\phi_i | \rightarrow \frac{v_i}{\sqrt{2}}$. so we apply the approximation $|\phi_i | = \frac{v_i}{\sqrt{2}}$ in (\ref{CosmicStringEOM}) and solve the equation to acquire
\begin{eqnarray}
    a(r)=\frac{n}{e} \left(1-D_0 r^{\frac{1}{2} }e^{-\omega r}\right), \label{a_Asymp_Result}
\end{eqnarray}
where $\frac{n}{e}=\frac{n_1}{Q_1}=\frac{n_2}{Q_2}$, and $\omega=\sqrt{2 \sum_{i=1,2}Q_i^2 v_i^2}$. This is the asymptotic solution of the $A^{\theta}(r)$ at the $r \rightarrow \infty$ limit. Here $D_0$ is a constant.

For the asymptotic behaviors of scalar fields at the $r \rightarrow \infty$ limit, substitute (\ref{a_Asymp_Result}) with (\ref{String_Field_Expansion}) into (\ref{CosmicStringEOM}), and preserve the linear terms of the $\frac{\partial V}{\partial \phi_i}$ only, we have
\begin{eqnarray}
    F^{\prime \prime}+ \frac{1}{r} F^{\prime}-\left[\frac{Q}{r} \left(a(r)-\frac{n}{e}\right)\right]^2 F \approx F^{\prime \prime}+ \frac{1}{r} F^{\prime}-\left[\frac{N}{r} \left(D_0 r^{\frac{1}{2} }e^{-\beta r}\right)\right]^2 F \approx M^2_{4 \times 4} F, \label{F_Equation}
\end{eqnarray}
where $F = [\text{Re}(f_1) - \frac{v_1}{\sqrt{2}}, \text{Im}(f_1), \text{Re}(f_2) - \frac{v_2}{\sqrt{2}}, \text{Im}(f_x) ]^T$ is the combination of the real scalar fields, $M_{4 \times 4}^2$ is the squared-mass matrix, as those in (\ref{EOM_Linear_Part_DomainWall}), and $Q=[Q_1,Q_2,]^T$, $N = \text{diag}[n_1, n_1, n_2,  n_2]$. Here, usually $N$ and $M_{4 \times 4}^2$ cannot be diagonalized at the same time, and since the  $\left(D_0 r^{\frac{1}{2} }e^{-\beta r}\right)$ term is suppressed by the exponent term, we further neglect this to reach the form
\begin{eqnarray}
    F^{\prime \prime}+ \frac{1}{r} F^{\prime} \approx M_{4 \times 4}^2 F
\end{eqnarray}
at the $r \rightarrow \infty$ limit. The asymptotic solution therefore is given by the form
\begin{eqnarray}
    \tilde{f}_i \sim \frac{e^{-m_i r}}{\sqrt{r}},
\end{eqnarray}
for $i=1,2,3$, where $m_i$ are the diagonalized masses in (\ref{Diag_M44}), and $\tilde{f}_i$ is the corresponding diagonalized field. For the Goldstone part $\tilde{f}_G$, the asymptotic solution is a constant $\tilde{f}_G = C_G$, indicating an additional rotation angle to the VEV for this to vanish. 

Due to the exponent term, the scalar fields converge much faster than the gauge field. Within (\ref{a_Asymp_Result}) the $D_0$ term also vanishes quickly compared with the $\frac{1}{r}$ behavior of the $A^{\theta}$. Therefore, for simplicity, we keep the $\propto 1/r$ term while neglecting all the other exponents in the asymptotic solution of the string fields, so the zero conditions for the root finder to find the reasonable $C_{1,2,3}$ are the distance of the numerical results and the asymptotic solutions at an enough far matching point $r = r_0$.

\subsection{Shooting Problems and the Target Function Constructions}

Practically, for the shooting method, two key problems lie at both the ``initial'' and ``final'' sides of the solution, when we try to match the numerical results with the approximated analytical asymptotic solutions.

For the initial guesses of the coefficients $C_{1,2}^+$ in (\ref{Asymp_Domain_Wall}) and the $C_{0,1,2}$ in (\ref{r0_Asymptot}), a random selection can hardly cause the target to converge, so we rely on our eyes to trace the orbits of the solutions in the field space, and adjust the initial coefficients manually to accomplish a better and better solution before the target is close enough for an automatic iteration is able to be started. Such a method is valid for some specific parameter points, however is impossible to be generalized to scan the parameter space. Therefore, after finding out the first solution manually, we then follow a ``stepwise'' strategy to traverse slowly into the target parameter point. The results of the previous point is input into the next point as the initial guesses.

For the ``target'' at the final sides, traditionally people find the root of a function dependent on the initial conditions, which is designed to vanish when initial values are correct. However, the equation solution algorithms usually fail to converge since the functions are extremely sensitive to the parameters. Therefore, alternatively we turn to the minimum finding algorithm for some ``target functions'' to accelerate the convergence. Such a ``target function'' incorporates various elements, including the ``distance'' between the two solutions, matching point coordinate, etc., to be addressed in detail in the following of this subsection.


\subsubsection{Domain Wall Target Functions}

For the CP-conserving potential described in (\ref{potential}), usually the profile of the cosmic string has a reflection symmetry according to its central plane. This can be realized if we rotate the real axis of both $\phi_{1,2}$ parallel to the central value of the profile, and the field values of the two corresponding symmetric points at both sides of the domain wall only differ by a complex conjugate. However, when CP violates, such a reflection symmetric plane disappears, we therefore define the ``highest point'' of the potential for the profile route to passes through as the ``central plane''. Regarding $z$ as the ``time parameter'', notice that (\ref{DomainWallEOM}) is similar to the Newton's second law equation of an object moving in a four-dimensional space. The highest potential energy in the orbit indicates the zero ``power'' point, in which $ \sum\limits_{t\in \lbrace1r, 1i,2r,2i \rbrace} \phi_t(z_0)'\phi_t(z_0)''=0$ should be satisfied. $z_0$, which is the position of the central plane, contributes to the target function as
\begin{eqnarray}
    W_0 z_0^2 \label{Central_Target}
\end{eqnarray}
to adjust the position of the domain wall to the target $z_0 = 0$. Here $W_c>0$ is the corresponding weight factor.



For the ``shooting'' processes, we start at some low enough $z_s<0$ where the initial conditions are set according to (\ref{Asymp_Domain_Wall}), in which $C^-_{1,2,3}=G_{0,1}=0$, and $C^+_{1,2,3}$ are the unknown parameters. For each of the point at $z>0$, define the squared ``distance''
\begin{eqnarray}
    D(z) = \text{min}_{\beta} \left\lbrace \sum_{i=1,2} \left| \phi_i(z) - \frac{v_{i \beta}}{\sqrt{2}} \right|^2 \right\rbrace, \label{Dz_Function}
\end{eqnarray}
between the current field value $\phi_i(z)$ and the target VEV defined by $v_{1 \beta} = v_1 e^{i \beta}$ and $v_{2 \beta} = v_2 e^{i \frac{\beta}{3}}$. Here $\text{min}_{\beta} \left\lbrace \dots \right\rbrace$ means to minimize the value as $\beta$ varies. Then we minimize the $D(z)$ for the second time dependent on $z$,
\begin{eqnarray}
    D(z_m) = \text{min}_z \left\lbrace D(z) \right\rbrace,
\end{eqnarray}
where $z_m$ is the coordinate for the minimum of $D(z)$. Then the corresponding part of the target function is design as
\begin{eqnarray}
    W_d D(z_m) - W_z z_m^2. \label{dz_Target}
\end{eqnarray}
$W_d>0$ and $W_z>0$ are the corresponding weights. Minimizing a combination of (\ref{Central_Target}) and (\ref{dz_Target}) results in a solution to hit the target vacuum as precisely as possible in a sufficient large $z$.

After that, we take into account the asymptotic solutions in (\ref{Asymp_Domain_Wall}). Practically (\ref{Central_Target})+(\ref{dz_Target}) finally cease to decrease due to the accumulating numerical errors. We therefore have to match the numerical solution with the asymptotic analytical evaluation for a more precise solution. 

At this stage, we again follow (\ref{Dz_Function}) to compute the $\beta_0$, which minimizes the squared ``distance'' at the matching point $z_0$. Then expand at the $z_0$, therefore $\phi_{1}(z_0) = \frac{v_1 e^{i \beta_0} + \phi_{1 r}(z_0) + \phi_{1 i}(z_0)}{\sqrt{2}}$, and $\phi_{2}(z_0) = \frac{v_2 e^{i \frac{\beta_0}{3}} + \phi_{2 r}(z_0) + \phi_{2 i}(z_0)}{\sqrt{2}}$ as in (\ref{phi12_Expansion}) before diagonalizing the mass matrix at the $\phi_{1r, 1i, 2r, 2i}(z_0)$ basis to rotate into $\tilde{\phi}_{1,2,3}(z_0)$ and $G(z_0)$ as in (\ref{shift to NGB direction}). By matching the values and derivatives with (\ref{Asymp_Domain_Wall}), we thereby solve the $C_{1,2,3}^{\pm}(z_0)$ and $G_0(z_0)$, $G_1(z_0)$ coefficients. A successful matching then requires the (nearly) vanishing $C_{1,2,3}^{+}(z_0)$ and $G_1(z_0)$. Then the relevant part of the target function is designed to be 
\begin{eqnarray}
    W_c \frac{\sum\limits_{i=1,2,3} \left[ C^{+}_i(z_0) \right]^2} {\sum\limits_{i=1,2,3} \left[ C^{-}_i(z_0) \right]^2} + W_G  \frac{ \left[ G_1(z_0) \right]^2}{\sum\limits_{i=1,2,3} {m_i^2 \left[ C_i^{2}(z_0) \right]^2}} - W_z z_0^2.
\end{eqnarray}
A small $G_0(z_0)$ might remain, and can be corrected by (\ref{delta_beta_correction}). Finally, all the $C^+_i(z_m)$ and $G_1(z_m)$ are neglected for a nearly smooth transition from the numerical solutions to the analytic asymptotic solutions.

During the whole processes, the programs might be aborted for us to adjust the weights $W_{0,d,z,c,G}$ by hand in order to accelerate the convergence. 

\subsubsection{Cosmic String Target Functions}

As we have addressed around (\ref{r0_Asymptot}), we start at a small $r = r_s > 0$ to estimate the field values to begin each shooting process. $C_{0,1,2}$ there are the initial parameters. For the $r \rightarrow \infty$ limit, we have to match with the asymptotic solution (\ref{a_Asymp_Result}) at some $r_m$, while neglecting all the exponent terms in $f_{1,2}$ and $a$, so it is appropriate for us to incorporate the ``distance'' between our numerical solutions and the asymptotic solutions in the target function
\begin{eqnarray}
    \sum_i {\left(\phi_i-\frac{v_i}{\sqrt{2}}\right)^2}+\left(A_{\theta}-\frac{n}{e r}\right)^2 . \label{Dist_String}
\end{eqnarray}
During each iteration at each ``shooting'', find the minimum of the (\ref{Dist_String}) at the $r_m$ to match with the asymptotic solution. Then minimize the following target function
\begin{eqnarray}
    W_d \left[ \sum_i {\left(\phi_i-\frac{v_i}{\sqrt{2}}\right)^2}+\left(A_{\theta}-\frac{n}{e r}\right)^2 \right] - W_r r_m^2
\end{eqnarray}
to find out the best solution satisfying the boundary conditions. $W_{d,r} > 0$ again are the weights for us to adjust. 

\subsection{Parameter Selections and Constraints}

In this paper, for convenience, we choose the $v_1$ and $v_2$ instead of the $\mu_1$ and $\mu_2$, together with the $\lambda_{1,2,3,12}$ as our free input parameters to evade the complicated processes solving the cubic equation set. The relationships between the two parameter sets are determined by the stationary point equations of the effective potential $\frac{\partial V}{\partial \phi_{1r}}=0$ and $\frac{\partial V}{\partial \phi_{2r}}=0$. $\frac{\partial V}{\partial \phi_{1i}}=0$ and $\frac{\partial V}{\partial \phi_{2i}}=0$ are automatically guaranteed by the invariance of the potential under the $C$-symmetry operation $\phi_{1,2i} \rightarrow -\phi_{1,2i}$. $\mu^{2}_{1,2}$ are then calculated to be,
\begin{eqnarray}
& & \mu^{2}_{1}=\left(2 \lambda_{1} v_{1}^{3}+\lambda_{12} v_{1} v_{2}^{2}-\lambda_{3} v_{2}^{3}\right)/ v_{1} ,\nonumber \\
& & \mu^{2}_{2}=\left(2 \lambda_{2} v_{2}^{3}+\lambda_{12} v_{1}^{2} v_{2}-3 \lambda_{3}  v_{1}v_{2}^{2}\right)/ v_{2}.
\end{eqnarray}


Finally, there remain the gauge coupling $Q_{1,2} g = Q_{1,2}$. We have to note that upper bounds exist for the U(1) gauge coupling constants\cite{Finger:1979yt,Ni:1982fi,Wang:1983vacuum}. Here, we define the ``fine-structure constant'' $\alpha_1$ as
\begin{eqnarray}
    \alpha_1=\frac{Q_1^2}{4 \pi},
\end{eqnarray}
which must satisfy $\alpha_1\lesssim 3$. The boundary conditions in (\ref{string boundary condition}) and the practical calculations tell us that in (\ref{Tensor_CosmicString}), the integrand term $\left|\frac{1}{r}\frac{\partial\phi_{i}}{\partial \theta}-i Q_{i} A^{\theta} \phi_{i}\right|^2$ is truncated at smaller $r$ as $\alpha_1$ increases, so the $\mu$ accordingly decrease. Actually (\ref{Life_Wall}) and (\ref{kappa_s}) also tell us that an appropriate lifetime $t_{\Gamma}$ requires a suppressed $\kappa_s \sim 1$, so $\mu$ must be controlled by a relatively larger $Q_1 g$. Therefore the bound $\alpha_1\lesssim 2.5$ becomes crucial.

Besides the gauge couplings, scalar couplings $\lambda_{1,2,3,13}$ are also bounded from the perturbative unitarity conditions\cite{Lee:1977eg,Lee:1977yc,Cynolter:2004cq,Aoki:2007ah} . Detailed discussions and calculations are beyond the scope of this paper, we therefore adopt a relatively safe $|\lambda_{1,2,3,13}| \lesssim \pi$. In this paper, Such constraints have been considered and the values of the $\alpha_1$, $\lambda_{1,2,3,13}$ are adopted within them.

\subsection{Scaling rules}

Since the shooting method takes enormous time, it is fairly difficult to proceed a full scan over the parameter space. Fortunately, in this subsection we will show the equivalence of the solutions among different parameter points by zooming the parameters in groups, thus eliminate the degrees of freedom by a value of two. The parameter space then becomes more compacted, and each benchmark point thereby represents a series of other equivalent points.

In Eq.~\eqref{general EOM}, in total 10 quantities are involved: $\{ \phi_1,\, \phi_2,\, A^{\mu},\, x^{\mu},\, \mu_1^2,\, \mu_2^2,\, \lambda_{1,2,3,12} \}$. They can be further classified into four groups:
\begin{itemize}
    \item fields, denoting $F= \{\phi_1, \phi_2, A^{\mu}\}$,
    \item dimensionless couplings, denoting $\lambda=\{\lambda_1, \lambda_2, \lambda_3, \lambda_{12}\}$,
    \item mass parameters, denoting $M^2 = \{\mu_1^2, \mu_2^2\}$,
    \item coordinates, denoting $x = \{x^\mu\}$.
\end{itemize}

Then we scale these quantities in groups.
\begin{itemize}
    \item $F \rightarrow {\gamma_1} F$, $M^2 \rightarrow {\gamma_1}^2 M^2$, $x \rightarrow \frac{1}{{\gamma_1}} x$, while $\lambda$ remains intact. 
    Here $\gamma_1$ is an arbitrary positive real number, and can be understood as a universal scaling factor for parameters with dimension(s). 
    Scale the quantities in Eq.~\eqref{general EOM} with these ${\gamma_1}$-dependent factors. 
    Notice that $\partial_{\mu} \rightarrow {\gamma_1} \partial_{\mu}$, then there will generate a universal factor ${\gamma_1}^3$ multiplied at both sides of the equations. All these solutions that can be connected with different ${\gamma_1}$ are equivalent.
    \item $\lambda \rightarrow {\gamma_2} \lambda$, $F \rightarrow \frac{F}{\sqrt{{\gamma_2}}}$, while $M^2$ and $x$ remain intact. Here ${\gamma_2}$ is an arbitrary positive real number, and can be understood as a scaling factor of the dimensionless parameters. 
    Scaling the quantities in Eq.~\eqref{general EOM} generates a universal factor of $\frac{1}{\sqrt{{\gamma_2}}}$ at both sides, connecting different parameter points that all $\lambda_{1,2,3,12}$ and $q^2$ are scaled as a group.
\end{itemize}

According to the discussions above, we can select the four ratios $\frac{\mu_2^2}{\mu_1^2}$, $\frac{\lambda_{2,3,12}}{\lambda_1}$ as the free parameters. Practically we choose
\begin{eqnarray}
    R_{12} = \frac{v_1}{v_2} \label{R12_Def}
\end{eqnarray}
in place of the $\frac{\mu_2^2}{\mu_1^2}$. It is also easy to calculate from (\ref{Tensor_Domainwall}) and (\ref{Tensor_CosmicString}) that under the transformations of the scaling factors ${\gamma_1}$ and ${\gamma_2}$, the scaling rules for the tenses $\sigma$ and $\mu$ become $\sigma \rightarrow \frac{{\gamma_1}^3}{{\gamma_2}} \sigma$, $\mu \rightarrow \frac{{\gamma_1}^2}{{\gamma_2}}$. Therefore $\kappa_s \propto \frac{\mu^3}{\sigma^2}  \rightarrow \frac{1}{{\gamma_2}} \kappa_s$.

\subsection{Parts of the Results: Phases and Profiles of the Walls and Strings}

As we have addressed in the previous sections, if $v_1 \gg v_2$, the whole breaking processes can be split in two steps, $U(1) \rightarrow Z_3$ followed by $Z_3 \rightarrow \lbrace e \rbrace$. Intuitively, when $v_1 \sim v_2$ or even $v_1 < v_2$, the meaning of the intermediate ``$Z_3$ symmetry'' gradually becomes ambiguous, and accordingly the ``$Z_3$-domain wall'' finally disappears. In this paper, we are particularly interested in the question that how the $Z_3$ symmetry as well as the corresponding domain wall disappear as $v_1$ drops.

\begin{figure}[t!]
\includegraphics[width=0.4\textwidth]{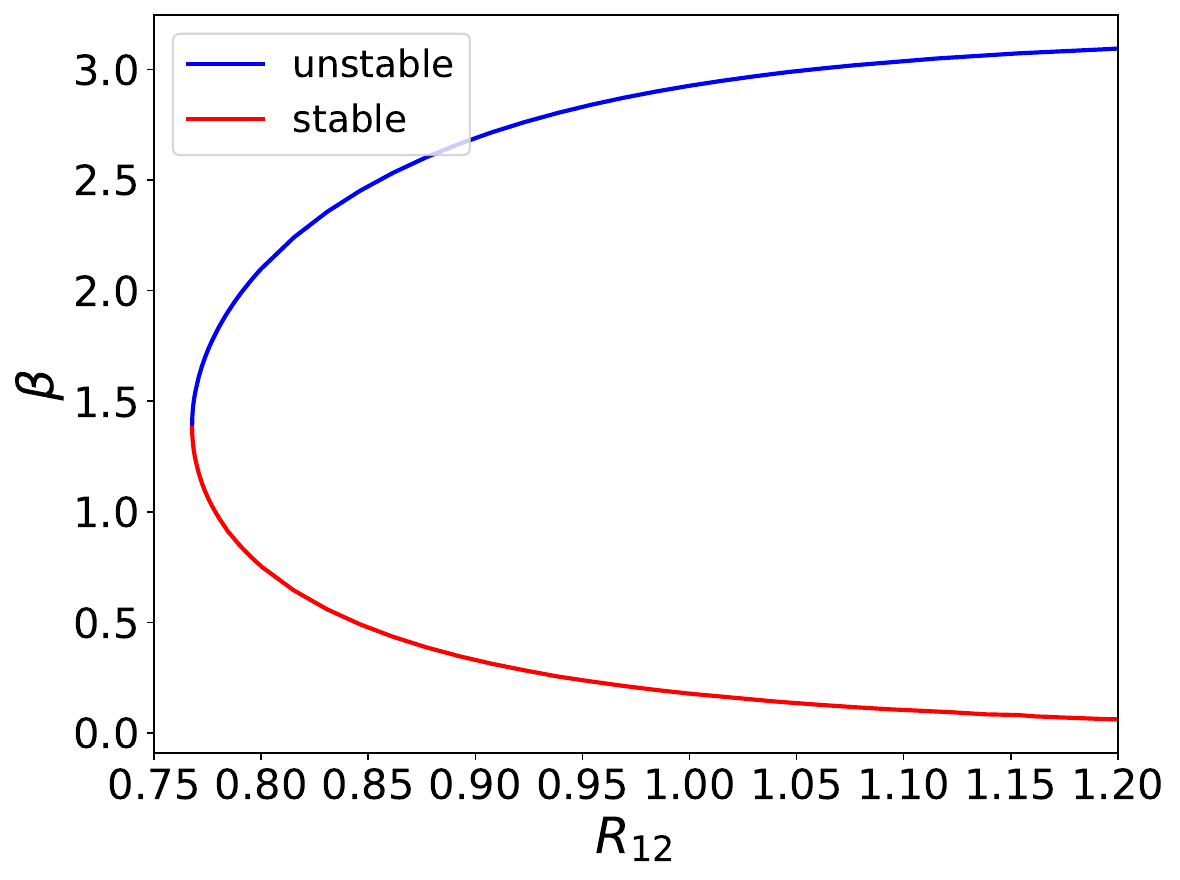}
\includegraphics[width=0.4\textwidth]{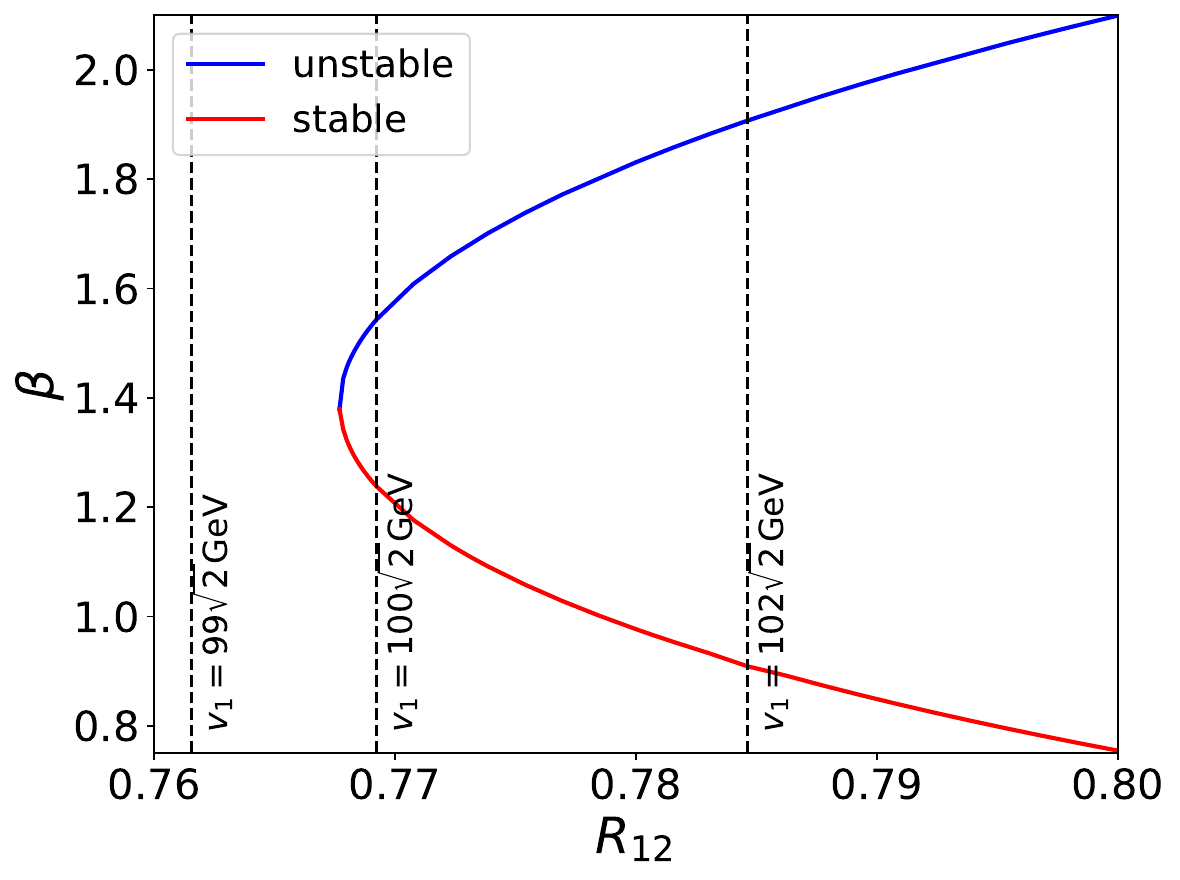}
\caption{\label{PhaseDiagram} phase diagrams of the stable and unstable domain wall phases. The right panel is a zoomed version around the critical point $R_{12} \approx 0.7677$ in which the two phases converge. The parameters adopted are $v_2=130\sqrt{2}\text{GeV}$, $\lambda_1=1$, $\lambda_2=1.6$, $\lambda_{12}=0$, and $\lambda_3=1$. $R_{12}$ is defined in (\ref{R12_Def}). The stability of the domain walls denoted in this figure will be illustrated in Fig.~\ref{SigmaAngle}. }
\end{figure}

To answer this question, we calculate the profiles of the domain walls by the shooting algorithm. At the left panel of the Fig.~\ref{PhaseDiagram} we show an example in which all other parameters are fixed except the $R_{1 2}$, where $R_{12}=\frac{v_1}{v_2}$ is the relative size of two complex VEVs. When $R_{1 2} \rightarrow \infty$, the bias angle $\beta$ defined in (\ref{DomainWallBoundaryCondition}) tend to vanish, recreating the standard $Z_3$ domain wall. On the contrary, when $R_{1 2}$ drops down, $\beta$ gradually ascends. When $R_{1 2} < 1$, the domain wall solution still exists for a while before the $\beta$ suddenly springs up $\rightarrow 1.378$, which is closed to $\frac{\pi}{2}$, and the curve then seems to veer backwards, forming another phase of the domain wall with the bias angle $\beta \gtrsim \frac{\pi}{2}$, which co-exists with the $\beta \rightarrow 0$ one. Therefore the problem lies at the stability of these two domain wall solutions.

\begin{figure}[t!]
\includegraphics[width=0.3\textwidth]{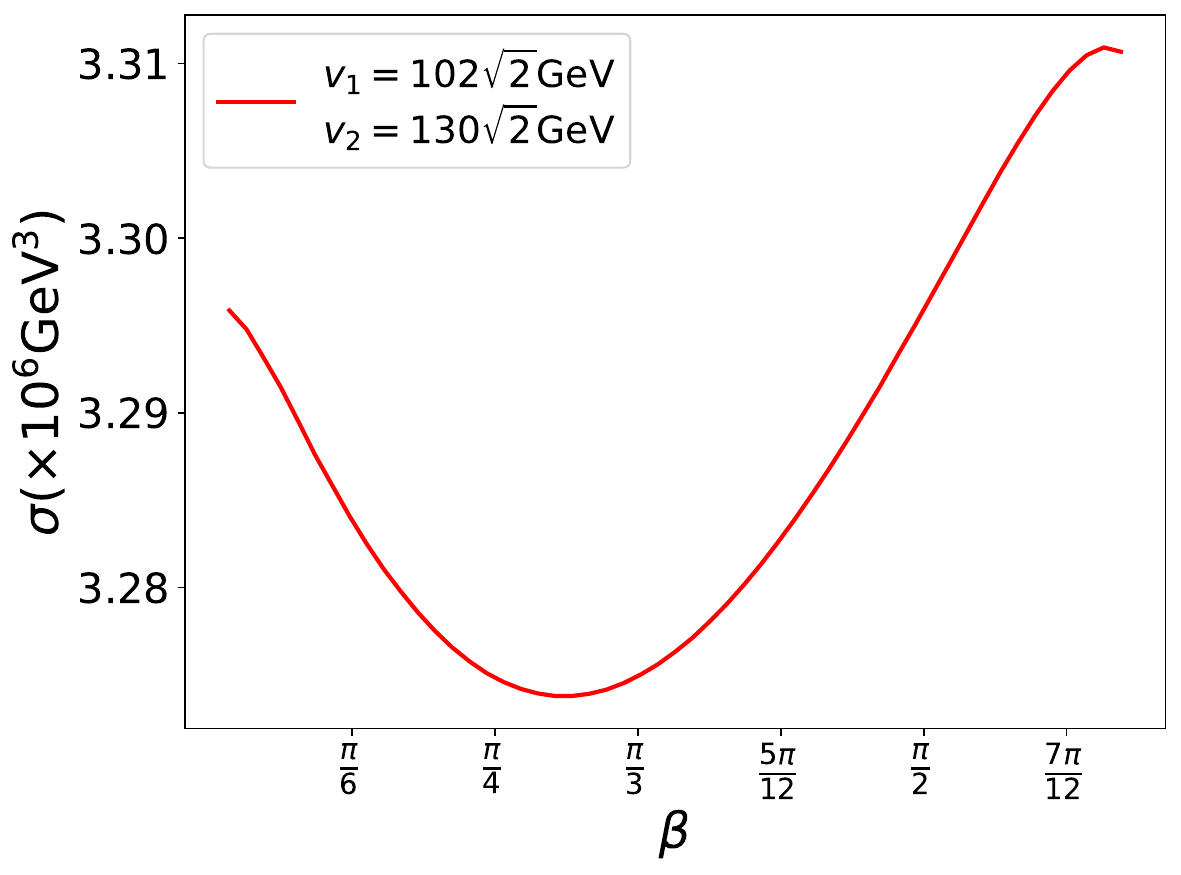}
\includegraphics[width=0.3\textwidth]{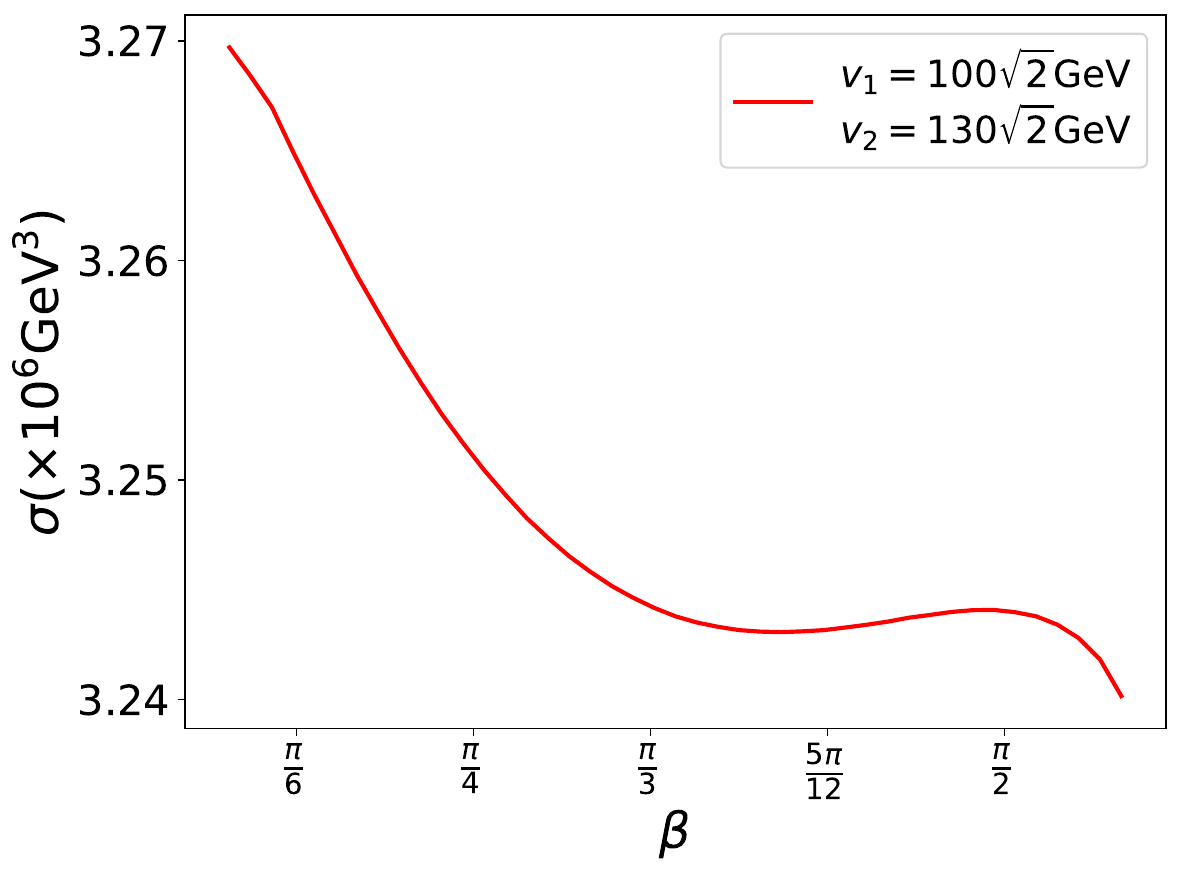}
\includegraphics[width=0.3\textwidth]{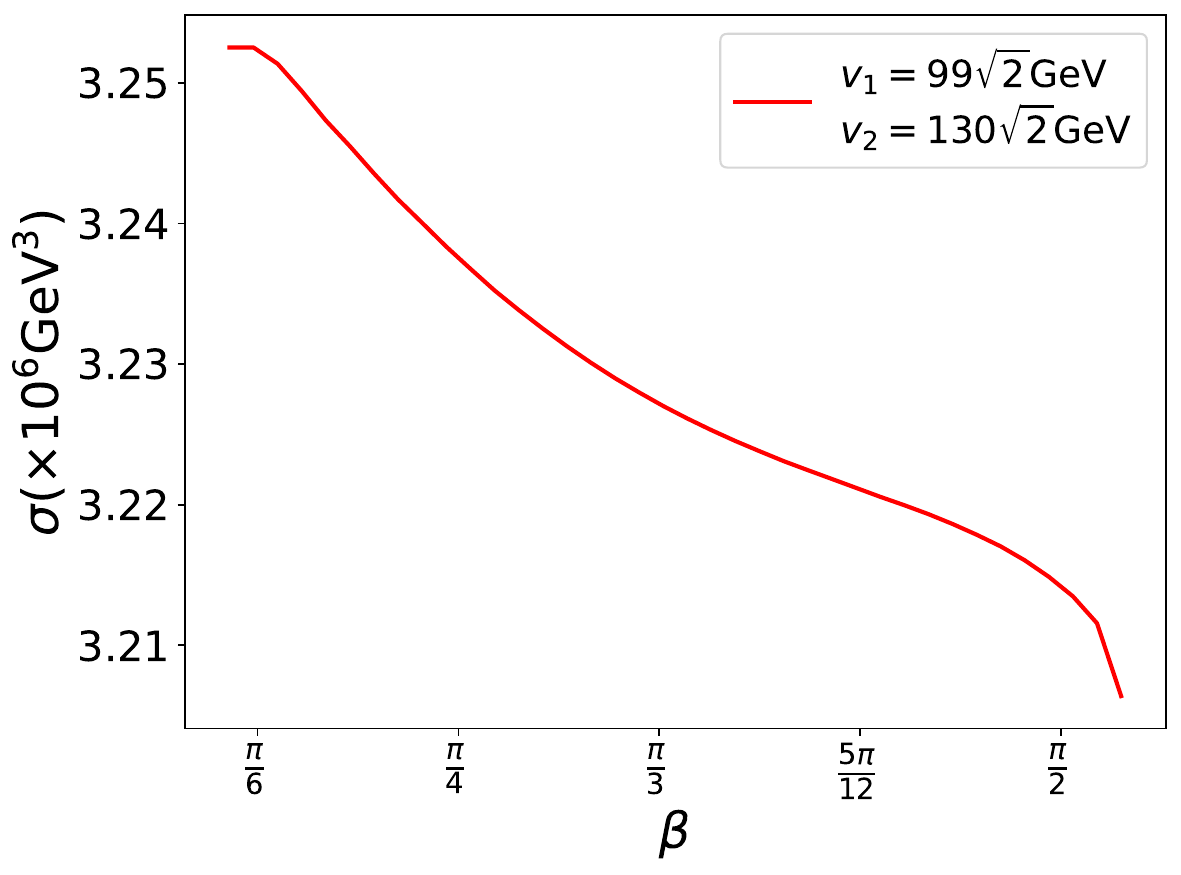}
\caption{\label{SigmaAngle} The conditional domain wall energy dependent on the bias angle $\beta$. The parameters are the same with Fig.~\ref{PhaseDiagram}, and $v_2$ varies. The extrema at the first two panels correspond to the intersections of the vertical straight lines with the curve in the right panel of Fig.~\ref{PhaseDiagram}. Minima correspond to the meta-stable structures, while maxima indicate the unstable ones. }

\end{figure}

The path deformation algorithm can help us solve this problem. With this algorithm, unlike the shooting-method, one can fix both the vacua at a particular $\beta$ for the domain wall to transit. Then different $\beta$'s are appointed and attempted, and for each of the $\beta$, a conditional minimum of the domain wall actions, or equivalently the domain wall tension $\sigma(\beta)$ defined in (\ref{Tensor_Domainwall}), is therefore acquired when the profile converges. In Fig.~\ref{SigmaAngle}, we plot the (conditional) tensions of the domain wall ``solutions'' as the functions of the bias angle $\beta$ for $v_1=102\sqrt{2}$, $100\sqrt{2}$, $99\sqrt{2}$ GeV respectively at the left, middle, and right panels. In the first three panels the extrema are accommodated. Comparing with the right panel of the Fig.~\ref{PhaseDiagram}, which is the zoomed version of the left panel around the ``veering point'', we see that both of the extrema in each $v_1$ are compatible with the two domain wall solutions acquired from the shooting algorithm with the same parameters (corresponding to the two intersection points of each $R_{12} = \frac{102,101,100}{130}$ vertical lines with the red and blue curve). The maximum in the curve, or more precisely, the ``saddle point'' in the full profile space in which $\beta > 1.378$ indicates the corresponding solution is unstable. For the $\beta$ values outside the presented range in Fig.~\ref{SigmaAngle}, the path deformation algorithm fail to converge into a simple domain-wall profile, maybe due to the reason that the simple ``wall-solutions'' corresponding to these $\beta$ values are no longer conditional minima of the $\sigma$ for the fixed $\beta$ values, which are required for the validity of this algorithm in these $\beta$ value conditions. Of course, only the meta-stable local minimum that $\beta < 1.378$ is preserved. 

\begin{figure}
    \centering
    \includegraphics[width=0.7\textwidth]{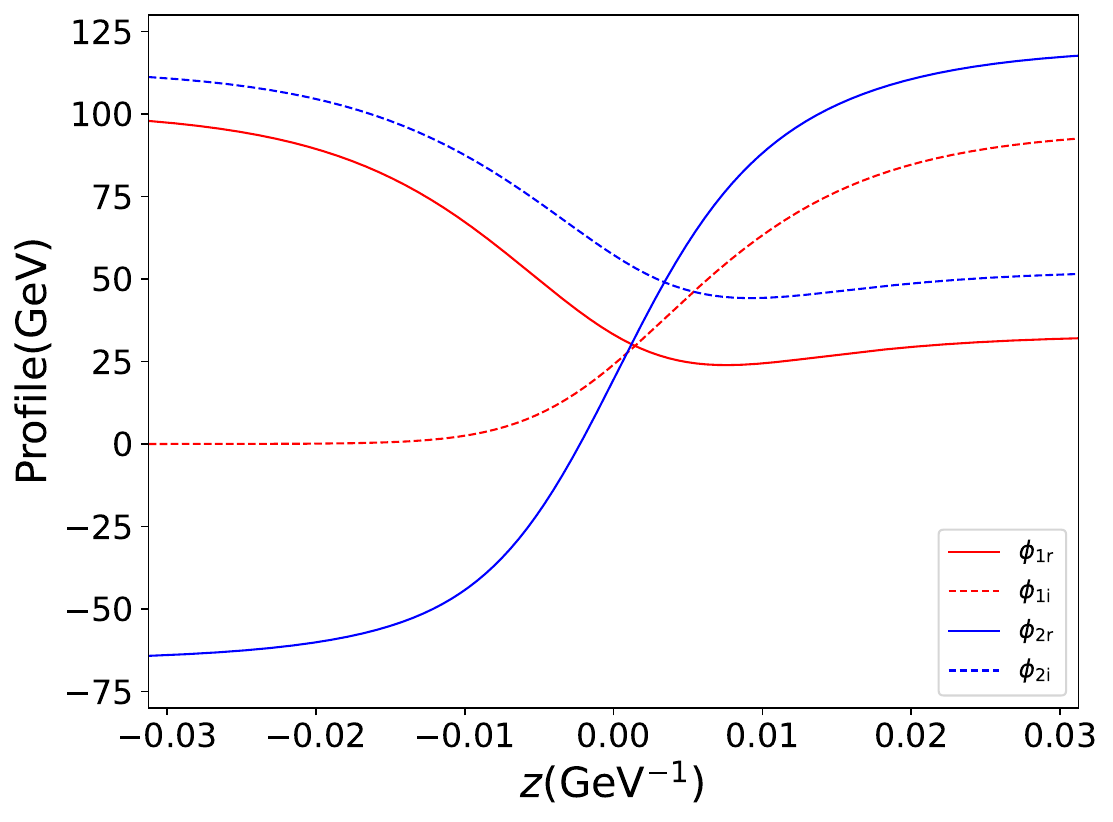}
    \caption{Profile of the domain wall for the benchmark point III in Tab.~\ref{BenchmarkStringEatingDW}.}
    \label{Profile_Fig}
\end{figure}

\begin{figure}
    \centering
    \includegraphics[width=0.49\textwidth]{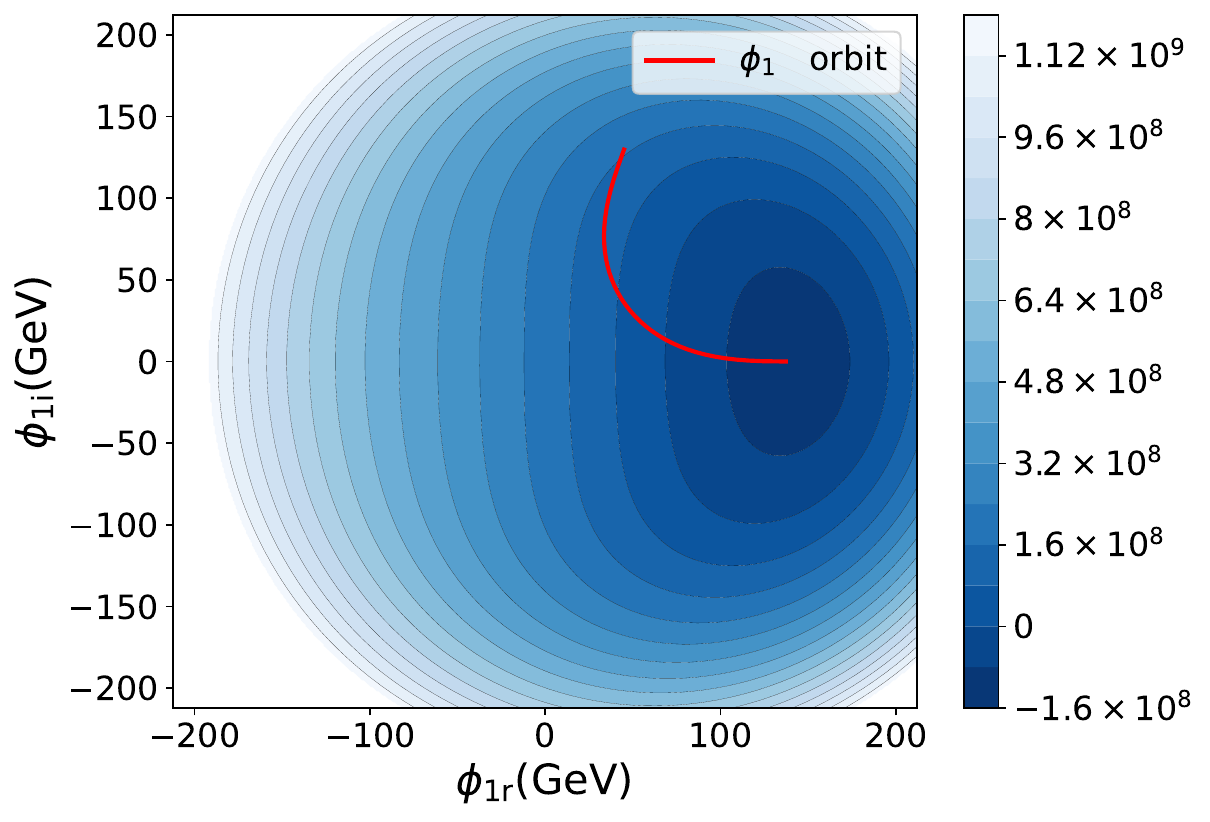}
    \includegraphics[width=0.49\textwidth]{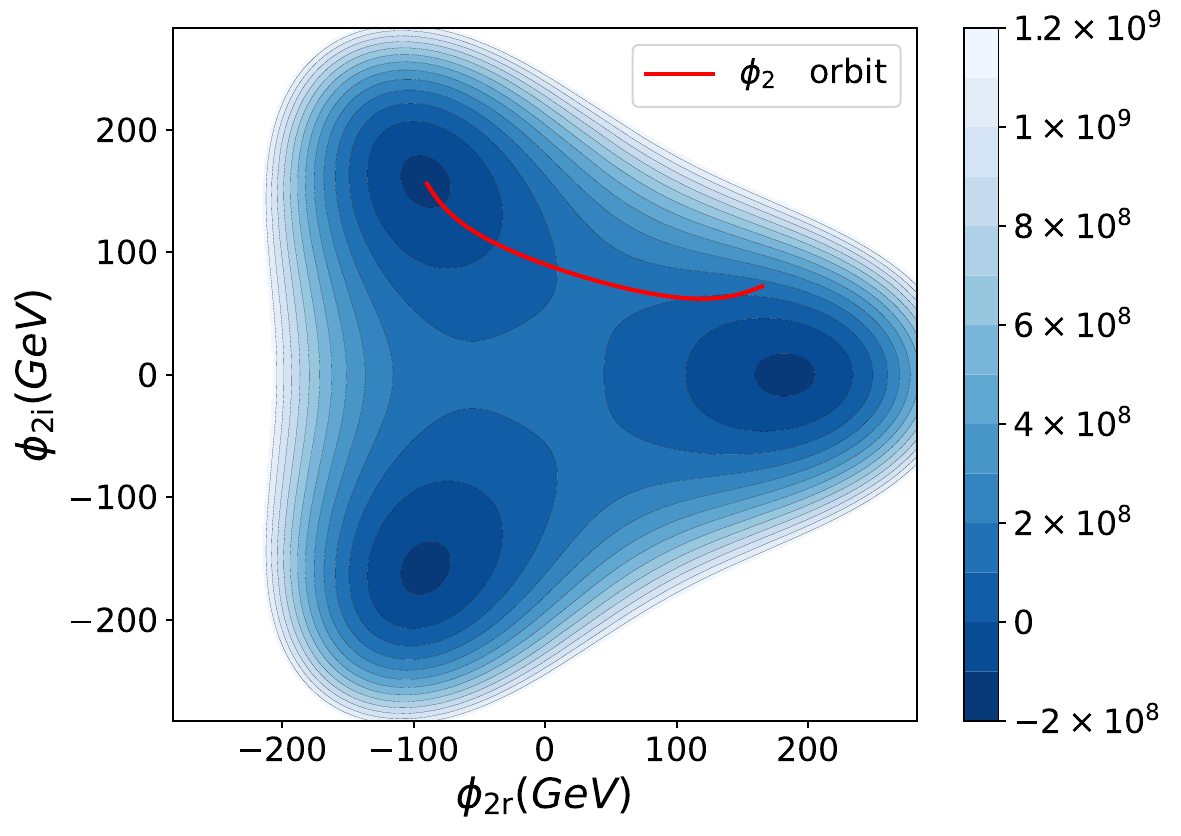}
    \caption{Orbit of the fields in the $\phi_{1,2}$ planes. Parameters are the same as Fig.~\ref{Profile_Fig}. The potential values are depicted in the conditions that $\phi_1$ or $\phi_2$ are fixed for the left and right panels, respectively. The argument difference of the beginning and ending points of the curve at the left panel indicates the bias angle, which is $\beta_1\approx 1.238$ in this figure. At the right panel a deviation from the $Z_3$ vacuums becomes obvious.}
    \label{OrbitPhi}
\end{figure}

As the $R_{1 2}$ further decreases, the two solutions seem to merge at $\beta = 1.378 \approx \frac{\pi}{2}$ before both of them annihilate. The right-bottom panel at Fig.~\ref{SigmaAngle} verifies the disappearance, in which all extrema vanish, validating our failure to find a domain wall solution by the shooting algorithm, as indicated by the $R_{12}=\frac{99}{130}$ vertical line labelled at the bottom-right panel of Fig.~\ref{PhaseDiagram}.

In Fig.~\ref{Profile_Fig}, we select a benchmark point to plot the domain wall profile. In Fig.~\ref{OrbitPhi} the orbits of the field configuration in the $\phi_{1,2}$ spaces are presented, above the background of the potential value at the $\langle \phi_{2} \rangle = \frac{v_2}{\sqrt{2}}$ and $\langle \phi_{1} \rangle = \frac{v_1}{\sqrt{2}}$ slices respectively. Here the bias angle $\beta = {1.238}$ deviating from the $Z_3$ symmetry significantly.

\begin{figure}
    \centering
    \includegraphics[width=0.4\textwidth]{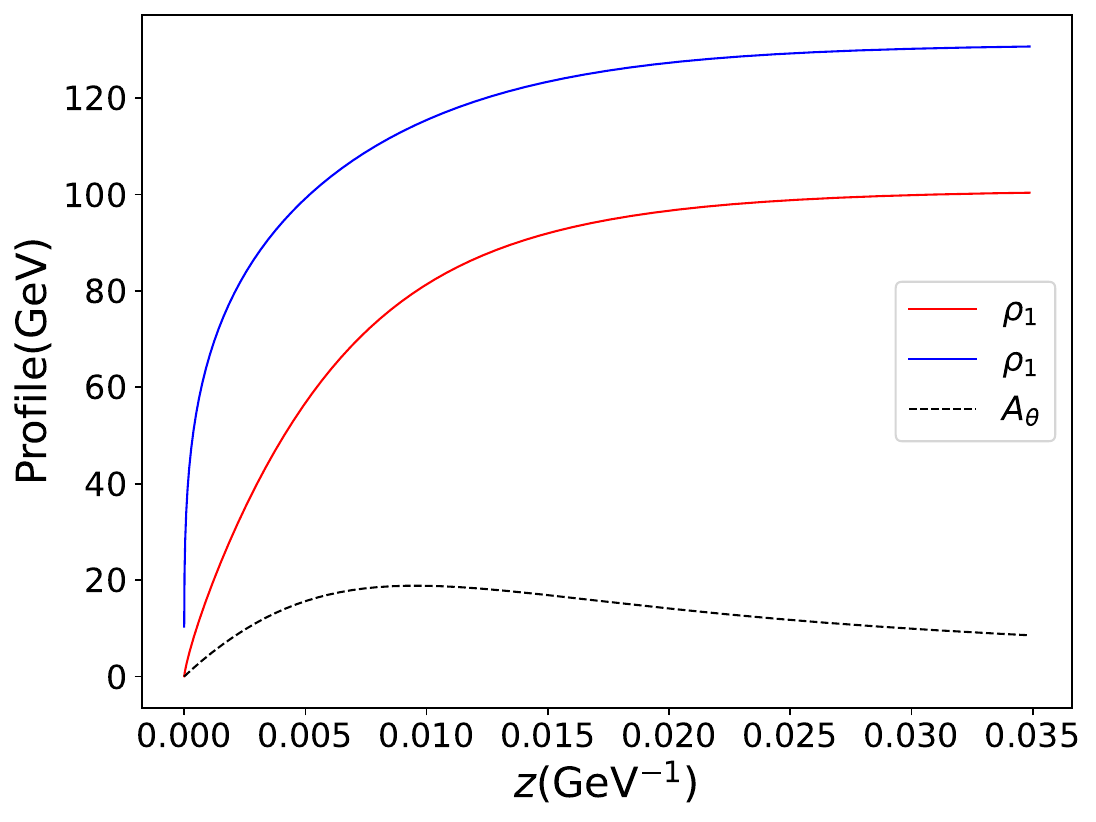}
    \caption{Profile of the string solution with $n_1$ given by (\ref{Fractional_n1}), in which $n=1$ and the bias angle $\beta = 1.238$ are adopted. Parameters are the same as Fig.~\ref{Profile_Fig}, charges of gauge fields are $Q_1=2.7$ and $Q_2=0.9$.}
    \label{StringProfile}
\end{figure}

Such a metastable domain wall finally disappears, either by surrounding the hole(s) or bounded by cosmic string(s). The corresponding cosmic string solution with a non-zero bias angle $\beta$ is presented in Fig.~\ref{StringProfile} as an example.

\section{Gravitational wave from hybrid defects}\label{GW}\label{part 5}


After the $U(1)$ symmetry is spontaneously broken, domain walls, cosmic strings, and the associated rims and conjunctions sketched in Fig.~\ref{sketch_rim} and Fig.~\ref{sketch_Conjunctions} can all form through the Kibble mechanism. This hybrid network is initially unstable and evolves dynamically until it settles into a scaling regime. A full dynamical simulation of this process is beyond the scope of this paper; instead, following Ref.~\cite{Dunsky:2021tih}, we estimate the resulting gravitational wave signal under a small set of simplifying assumptions, outlined below.

When $v_1 \approx v_2$, the two scalar fields acquire their VEVs during a single phase transition, so the wall-string network forms all at once. Immediately afterwards, rims are pulled toward nearby rims and conjunctions by the tension of the domain walls they bound, and fuse with them; Fig.~\ref{merge_3DW} and Fig.~\ref{merge_2DW} illustrate two such merging processes. If enough defects form before the network reaches scaling, the walls come to dominate the energy budget over the strings, owing to their larger scaling index. What happens next depends on the relative abundance of rims:
\begin{enumerate}
    \item if rims are rare compared with conjunctions, the merging process exhausts the available rims quickly, leaving behind a network of conjunctions. The surviving domain walls then decay only by nucleating new cosmic strings on their surface --- the ``strings eating walls'' scenario of Ref.~\cite{Dunsky:2021tih};
    \item if rims are instead abundant, they consume essentially all of the walls and conjunctions before the walls can dominate, leaving a pure string network --- the ``walls eating strings'' scenario of Ref.~\cite{Dunsky:2021tih}.
\end{enumerate}
A third possibility is that a stable domain wall never forms at all, depending on the details of the symmetry-breaking history; this includes the case where every wall created before the end of the scaling regime shrinks directly into strings via the processes of Fig.~\ref{merge_3DW}, Fig.~\ref{merge_2DW}, and Fig.~\ref{merge_Rims}. In either of these two cases, only cosmic strings remain, radiating gravitational waves as described in Ref.~\cite{Sousa:2013aaa}.

Strictly speaking, the surviving ``strings'' -- rims included -- are a mixture of configurations with different winding numbers $n_1$ defined in Eq.~\eqref{String_Field_Expansion}, and the relative abundance of each configuration would affect the resulting spectrum. Estimating these abundances is beyond our present simulation capability, so as a preliminary approximation we assume that the lowest allowed $n_1$ dominates each mixture.

Finally, in the hierarchical limit $v_1 \gg v_2$, the symmetry breaks in two steps, as discussed in the earlier literature \cite{Kibble:1982dd,Dunsky:2021tih}; we include this case for a comparison. The rest of this section works through the gravitational wave calculation for each of these scenarios in turn.

\subsection{Strings Eating Domain Walls} \label{StringEatingWallGW}
In Ref.~\cite{Dunsky:2021tih}, the ``strings eating walls'' scenario arises when no cosmic string network pre-exists --- for instance, because an epoch of inflation has diluted it away --- so that the domain walls collapse purely through quantum tunnelling, nucleating rims at the edges of the holes eaten into them. In our model, the wall-string network is instead already present when the rims begin to nucleate, but we can still follow the steps of Ref.~\cite{Dunsky:2021tih} to estimate the gravitational wave spectrum in this regime, treating the domain walls as the dominant source. Adopting the ``velocity one-scale'' ansatz \cite{Sousa:2013aaa}, in which the network maintains roughly one domain per Hubble volume, the walls begin nucleating strings at a time \cite{Kibble:1982dd,Preskill:1992ck}
\begin{eqnarray}
    t_{\Gamma} \sim \frac{1}{\sigma^{\frac{1}{3}}} \mathrm{e}^{\frac{16 \pi \kappa_s}{9}}, \label{Life_Wall}
\end{eqnarray}
where
\begin{eqnarray}
    \kappa_s = \frac{\mu^3}{\sigma^2}. \label{kappa_s}
\end{eqnarray}
Following the pure-domain-wall treatment of Ref.~\cite{Hiramatsu:2013qaa}, the resulting gravitational wave spectrum is estimated to be
\begin{eqnarray}
    \Omega_{\text{GW}}(f) \approx \Omega_{\text{GW, max}} \left\lbrace \begin{array}{cl}
        \left(\frac{f}{f_{\text{peak}}} \right)^{-1}, & f>f_{\text{peak}} \\
        \left(\frac{f}{f_{\text{peak}}} \right)^{3}, & f \leq f_{\text{peak}}
    \end{array} \right. ,
\end{eqnarray}
peaked at the frequency
\begin{eqnarray}
    f_{\text{peak}} \approx \frac{1}{t_{\Gamma}} \frac{a(t_{\Gamma})}{a(t_0)},
\end{eqnarray}
where $a(t)$ is the scale factor of the expanding universe. Assuming that most of the walls decay promptly at $t = t_{\Gamma}$, the peak amplitude is
\begin{eqnarray}
    \Omega_{\text{GW, max}} \approx \frac{16 \pi}{3} [(G \sigma t_{\Gamma})^2 + 2 x G \sigma t_{\Gamma}] \Omega_r \left( \frac{g_{*0}}{g_*(t_{\Gamma})} \right)^{\frac{1}{3}},
\end{eqnarray}
where $\Omega_\mathrm{r} = 9.038 \times 10^{-5}$ is the radiation energy density today, $g_*(t)$ is the effective number of relativistic degrees of freedom at time $t$, $g_{*0}$ its value today, and $x$ is the efficiency with which the walls' energy is converted into gravitational radiation as the strings nucleate. We take $x=1$ throughout this paper.

\subsection{Domain Walls Eating Strings}

Ref.~\cite{Dunsky:2021tih} also considers the opposite situation, ``walls eating strings'', in which loops of cosmic string are already present before domain walls form inside them. In our model this scenario is realised whenever rims dominate the string content of the wall-string network, closely mirroring the setup of Ref.~\cite{Dunsky:2021tih}; we therefore adapt its steps directly to compute the gravitational wave spectrum.

A domain wall nucleates inside a pre-existing string loop at time $t_k$, with initial size $l_k = \alpha t_k$. Simulations typically give $\alpha \approx 0.1$\cite{Blanco-Pillado:2017oxo,Blanco-Pillado:2013qja}, being the ratio of the loop formation length to the horizon size. The wall then shrinks the loop, so its size $\tilde{l}$ decreases with the time according to
\begin{eqnarray}
    G \mu (\tilde{t}-t_k) = \int_{\tilde{l}}^{\alpha t_k} dl^{\prime} \frac{1+\frac{l^{\prime}}{2 \pi R_c}}{\Gamma(l^{\prime})}, \label{tk_Equation}
\end{eqnarray}
where
\begin{eqnarray}
    R_c = \frac{\mu}{\sigma}, \label{Rc_Def}
\end{eqnarray}
is the critical radius at which the wall and string tensions balance, and $\Gamma(l)$ is the reduced gravitational wave power\cite{Dunsky:2021tih}. A loop of length $\tilde{l}$ at time $\tilde{t}$ then radiates the spectrum
\begin{eqnarray}
    \frac{dP(\tilde{l}, \tilde{t})}{d\tilde{f}} = \Gamma(\tilde{l}) G \mu^2 l g(\tilde{f}\tilde{l}), \label{Spectrum_P}
\end{eqnarray}
where $\tilde{f}$ is the emitted frequency and $g(x)$ is the discrete spectrum
\begin{eqnarray}
    g(x) = \sum_n \mathcal{P}_n \delta (x-\xi n)\,, \label{Spectrum_g}
\end{eqnarray}
with $\xi$ relating to the loop length to its oscillation period.
Since the domain wall and string energy densities are comparable in this work, we take $\xi = \pi$. This spectrum redshifts as $\tilde{f} \rightarrow f = \tilde{f} a(\tilde{t})/a(t)$ while the universe expands. Summing over the contributions of every loop, together with the ones of the domain walls bounded by loops, the gravitational wave spectrum observed today is
\begin{eqnarray}
    &\Omega_{\text{GW}} = &\sum_n \frac{8 \pi (G \mu)^2}{3 H_0^2} \int_{t_{\text{sc}}}^{t_0} d\tilde{t} \frac{a(\tilde{t})^5}{a(t_0)^5} \left( \frac{\mathcal{F} C_{\text{eff}}(t_{k,n})}{\alpha t_{k{\color{red} ,n}}^4} \frac{a(t_{k,n})^3}{a(\tilde{t})^3} \right) \nonumber \\
    & & \times \mathcal{P}_n \frac{\xi n}{f} \left( 1+ \frac{\xi n}{2 \pi R_c f} \frac{a(\tilde{t})}{a(t_0)} \right) \frac{\Gamma(\alpha t_{k{\color{red} ,n}}) \theta(t_* - t_{k,n})}{\Gamma(\alpha t_{k,n}) G \mu + \alpha \left( 1 + \frac{\alpha t_{k,n}}{2 \pi R_c} \right)}, \label{WallEatString_Result}
\end{eqnarray}
where $t_{k,n}$ solves Eq.~\eqref{tk_Equation} for each $\tilde{t}$, with $\tilde{l}$ replaced by $\tilde{l}_n$ --- the $n$-th nonzero point of the $\delta$-functions in Eqs.~\eqref{Spectrum_P} and \eqref{Spectrum_g}. The lower integration limit $t_{\text{sc}}$ marks the onset of the scaling regime. $t_{*}=\text{max}(R_c, t_{\text{DW}})$ is the time network collapsing, where $t_{\text{DW}}$ is the wall formation time. Finally, at each $\tilde{t}$ the corresponding source frequency is $\tilde{f} = f a(t_0)/a(\tilde{t})$, with $t_0$ the present time.


\subsection{Pure String Scenario}

As mentioned, it is also possible that only cosmic strings form, with no surviving domain walls. In this ``no wall'' scenario we follow the Ref.~\cite{Vilenkin:2000jqa,Cui:2018rwi} to compute the spectrum radiated by a network of pure string loops, defined as the ratio of the gravitational wave energy density per logarithmic frequency interval to the critical density
\begin{eqnarray}
    \Omega_{\text{GW}}=\frac{1}{\rho_{\text{crit}}}\frac{d \rho_{\text{GW}}}{d\text{log}f}.
\end{eqnarray}
At the time $t$, the gravitational wae spectrum at frequency $f$ is given by
\begin{eqnarray}
    \frac{d\rho_{\text{GW}}(t)}{df}= \int^{t}_{t_{sc} }d \tilde{t} \left(\frac{a(\tilde{t})}{a(t)}\right)^4 \int^{l}_{0} d \tilde{l} \frac{dn\left(\tilde{l},\tilde{t}\right)}{d\tilde{l} }\frac{dP\left(\tilde{l},\tilde{t}\right)}{d \tilde{f}}\frac{d\tilde{f}}{df},
\end{eqnarray}
where $dP/d\tilde{f}$ is again given by Eq.~\eqref{Spectrum_P} and $f = \tilde{f} a(\tilde{t})/a(t)$. The number density production rate of the loops, $dn/dl'$, is
\begin{eqnarray}
    \frac{dn\left(l^{'},t^{'}\right)}{dl^{'} }=\frac{1}{\alpha +\Gamma G \mu}\frac{\mathcal{F} C_{\text{eff}}(t_{k,n})}{\alpha t_{k,n}^4 }\frac{a(t_{k,n})^3}{a(\tilde{t})^3}. \label{PureString_Spectrum}
\end{eqnarray}
As a consistency check, taking $R_c \rightarrow \infty$ in Eq.~\eqref{WallEatString_Result} reproduces Eq.~\eqref{PureString_Spectrum} exactly.
{This is equivalent to the null wall-tension limit, in which the $\sigma$ vanishes in (\ref{Rc_Def}).}





\subsection{Parts of the Numerical Results for the Gravitational Wave Spectrum}

We now present numerical results for the gravitational wave spectra in each of the scenarios discussed above. In the $v_1 \sim v_2$ region of parameter space in which we mainly focus on, the fields $\phi_{1,2}$ acquire their VEVs simultaneously, and the three possible outcomes, ``strings eating walls'', ``walls eating strings'', and ``no-wall'', can all occur. 
For a comparison, we also show results in the hierarchical limit $v_1 \gg v_2$, where the symmetry breaks in the usual two-step scenario.

\begin{table}[b]

\begin{ruledtabular}
\begin{tabular}{lccccccc}
\textrm{name}&
\textrm{$v_1(\text{GeV})$}&
\textrm{$v_2(\text{GeV})$}&
\textrm{$\lambda_{1}$}&
\textrm{$\lambda_{2}$}&
\textrm{$\lambda_{12}$}&
\textrm{$\lambda_{3}$}&
\textrm{$ Q_1$}\\
\colrule
Point I & $100\sqrt{2}$ & $130\sqrt{2}$ & $\frac{5\pi}{8}$ & $\frac{15\pi}{16}$ & 0 & $\frac{5\pi}{8}$ & $\frac{27}{4}\sqrt{\frac{\pi}{10}}$\\
Point II& $\sqrt{2}\times 10^5$ & $1.3\sqrt{2}\times 10^5$ & $\frac{5\pi}{8}$ & $\frac{15\pi}{16}$ & 0 & $\frac{5\pi}{8}$ & $\frac{27}{4}\sqrt{\frac{\pi}{10}}$\\
Point III& $100\sqrt{2}$ & $130\sqrt{2}$ & $\frac{5\pi}{8}$ & $\pi$ & 0 & $\frac{5\pi}{8}$ & $\frac{27}{4}\sqrt{\frac{\pi}{10}}$\\
Point IV& $\sqrt{2}\times 10^5$ & $1.3\sqrt{2}\times 10^5$ & $\frac{5\pi}{8}$ & $\pi$ & 0 & $\frac{5\pi}{8}$ & $\frac{27}{4}\sqrt{\frac{\pi}{10}}$\\
Point V& $\sqrt{2}\times 10^{12}$ & $1.3\sqrt{2}\times 10^{12}$ & $\frac{5\pi}{8}$ & $\frac{15\pi}{16}$ & 0 & $\frac{5\pi}{8}$ & $\frac{27}{4}\sqrt{\frac{\pi}{10}}$\\
\end{tabular}
\end{ruledtabular}

\caption{
Benchmark parameter points used for the gravitational wave spectra of the ``strings eating walls'' scenario. 
}
\label{BenchmarkStringEatingDW}

\end{table}

\begin{figure}[t!]
\includegraphics[width=0.48\textwidth]{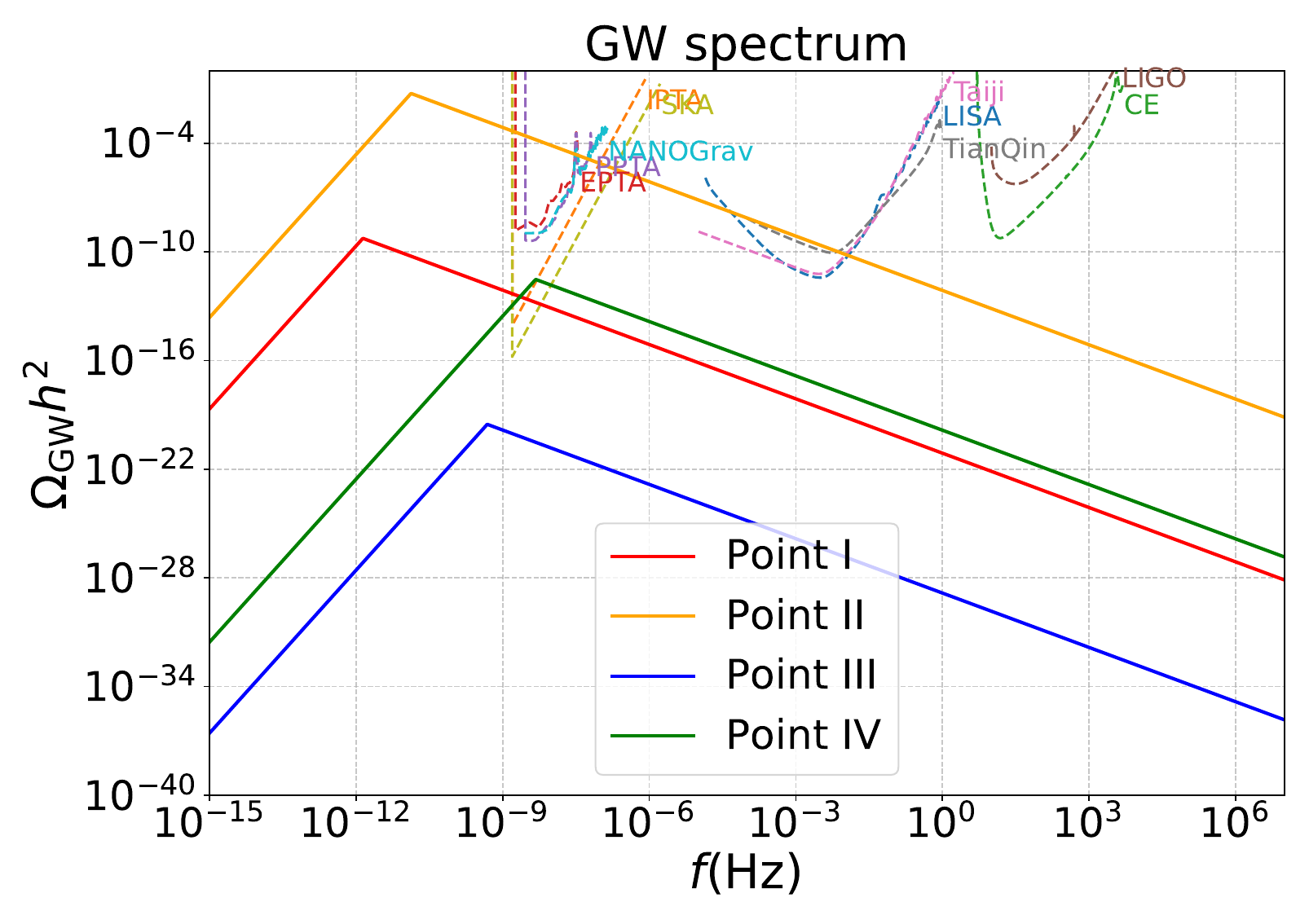}
\includegraphics[width=0.48\linewidth]{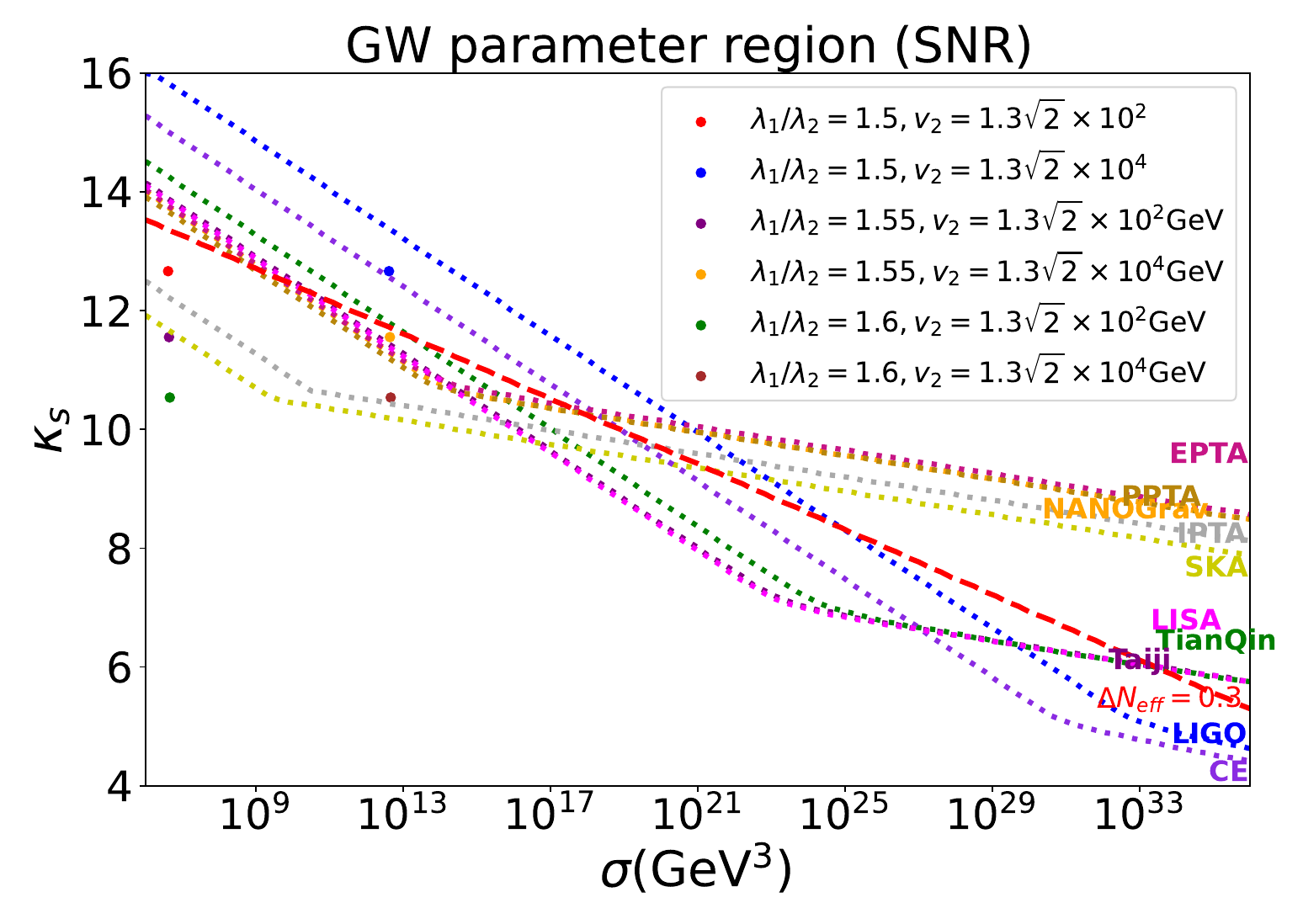}
\caption{The ``strings eating walls'' scenario for $v_1 \sim v_2$. Left: the gravitational wave spectra for the benchmark points of Tab.~\ref{BenchmarkStringEatingDW}, together with the sensitivities of several proposed gravitational wave experiments. Right: sensitivity curves of these experiments in the $\sigma$-$\mu$ plane, along with the same benchmark points; each curve corresponds to $\text{SNR}=1$ as defined in Eq.~\eqref{SNR_Def}, and parameter values above a given curve are more likely to be detected.}
\label{StringEatingWallSpectrum}
\end{figure}

For the ``strings eating walls'' scenario at $v_1 \sim v_2$, we use the benchmark points of Tab.~\ref{BenchmarkStringEatingDW} to plot the stochastic gravitational wave spectrum in the left panel of Fig.~\ref{StringEatingWallSpectrum}, together with the projected sensitivities of LIGO \cite{KAGRA:2013rdx}, CE \cite{LIGOScientific:2016wof}, LISA \cite{LISA:2017pwj}, Taiji \cite{Ruan:2018tsw}, TianQin \cite{Liang:2021bde}, SKA \cite{Janssen:2014dka}, NANOGrav \cite{NANOGRAV:2018hou}, PPTA \cite{Shannon:2015ect}, IPTA \cite{Hobbs:2009yy}, and EPTA \cite{EPTA:2015qep}.

Being the stochastic relics rather than transient signals, the background can in principle be extracted by extending the integration time. From Subsec.~\ref{StringEatingWallGW}, the spectrum depends only on the domain wall tension $\sigma$ and the string tension $\mu$, so we compute the signal-to-noise ratio (SNR) for each $(\sigma,\mu)$ pair using the sensitivity curves compiled in Refs.~\cite{Allen:1996vm,Allen:1997ad,Maggiore:1999vm,Thrane:2013oya,Schmitz:2020syl}
\begin{eqnarray}
    \text{SNR}=\sqrt{T \int_{f_{\text{min}}}^{f_{\text{max}}} df \frac{\Omega_{\text{GW}}(f) h^2}{\Omega_{\text{detector}}(f) h^2}}, \label{SNR_Def}
\end{eqnarray}
where $T$ is the observation time. In this paper take the assumption $T=4~\text{yrs}$ for every detector for simplicity. $\Omega_{\text{detector}}(f)$ is the corresponding detector sensitivity. The right panel of Fig.~\ref{StringEatingWallSpectrum} shows these benchmark points against the $\text{SNR}=1$ sensitivity curves of the proposed experiments.

The stochastic background also contributes an effective number of extra neutrino species $\Delta N_{\text{eff}}$, which is bounded by cosmological observations\cite{Planck:2018vyg,AtacamaCosmologyTelescope:2025nti,SPT-3G:2025bzu,Elbers:2025vlz}. We show the resulting $\Delta N_{\text{eff}} = 0.3$ bound as an additional constraint in the right panel of Fig.~\ref{StringEatingWallSpectrum}; since the bounds corresponding to any value of $\Delta N_{\text{eff}}$ between $0.03$ and $0.3$ nearly coincide on this plot, we display only this representative value rather than the full range. $\Delta N_{\text{eff}}$ is computed as
\begin{eqnarray}
    \Delta N_{\text{eff}}= \frac{8}{7}\left(\frac{11}{4}\right)^{\frac{4}{3}}\frac{\Omega_{\text{GW}}^0}{\Omega_{\gamma}},
\end{eqnarray}
where $\Omega_{\gamma}=2.47 \times 10^{-5}/ h^2$ and
\begin{eqnarray}
    \Omega_{\text{GW}}^0=\int \frac{df}{f} \Omega_{\text{GW}}(f).
\end{eqnarray}


\begin{figure}[t!]
    \centering
    \includegraphics[width=0.8\linewidth]{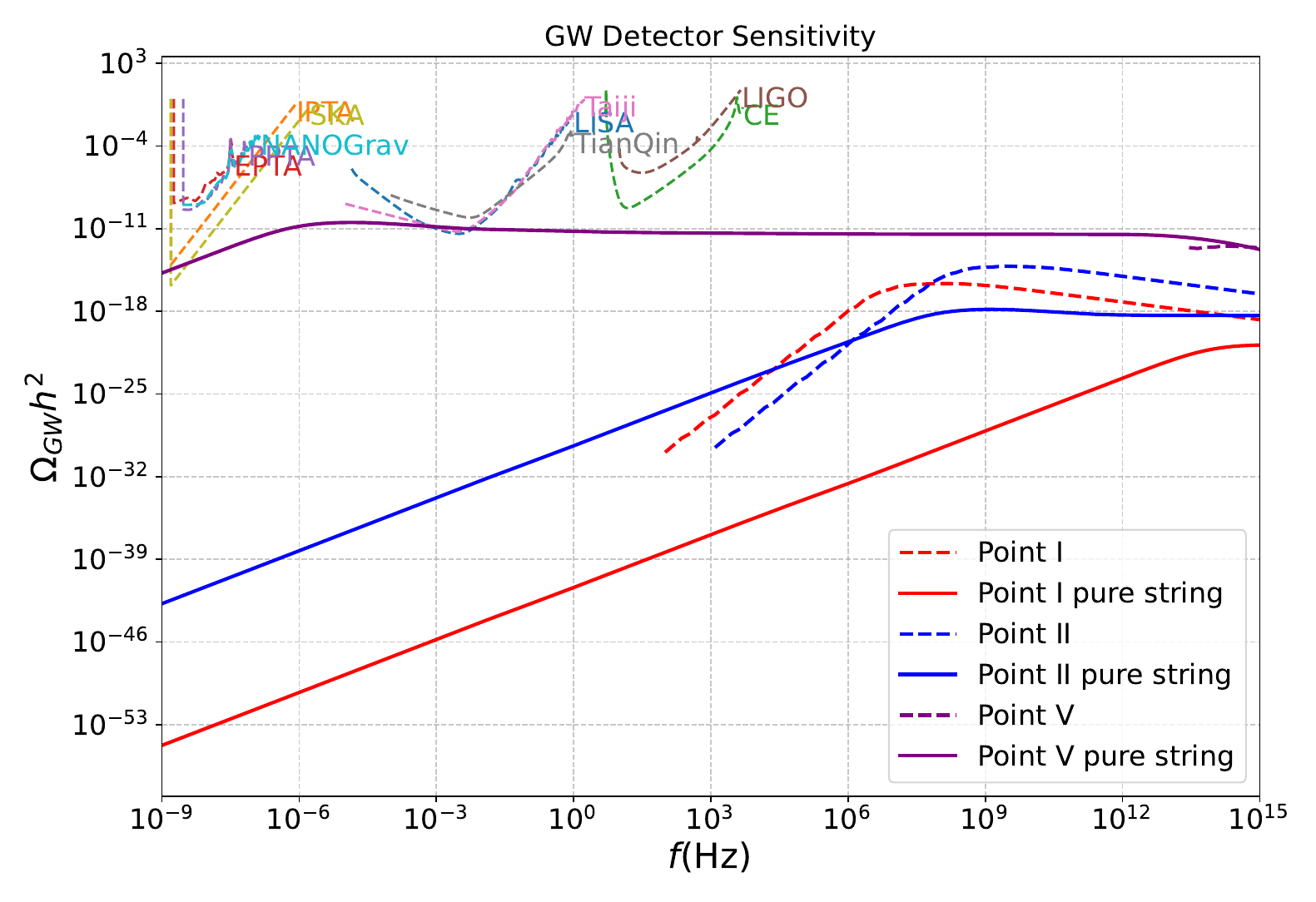}
    \caption{The ``walls eating strings'' scenario spectrum (dashed lines), alongside the ``no wall'' scenario spectrum (solid lines). The frequency axis extends to unrealistically high values, corresponding to defect formation temperatures far above the VEV scales; we nonetheless show the full curves, since the true high-frequency cutoff depends on details of defect formation beyond the scope of this estimate.}
    \label{StringBoundingWallSpectrum}
\end{figure}

For the ``walls eating strings'' scenario at $v_1 \sim v_2$, Fig.~\ref{StringBoundingWallSpectrum} shows the resulting spectrum (dashed lines) together with the detector sensitivity curves, again using the benchmark points of Tab.~\ref{BenchmarkStringEatingDW}. These frequencies lie well beyond the reach of any proposed detector.
Moreover, Eq.~\eqref{WallEatString_Result} shows that a nonzero spectrum requires the scaling regime to begin so early that the corresponding temperature $T_{\text{sc}} \propto \sqrt{M_{\text{pl}}/t_\text{sc}} \gg v_{1,2}$, i.e., the symmetry would have to break at a temperature far above its own breaking scale, which is physically implausible. The ``walls eating strings'' curves in Fig.~\ref{StringBoundingWallSpectrum} are therefore not physically realizable, though we show them for completeness.
As a comparison, the same figure shows the ``no wall'' scenario spectrum (solid lines) for the identical benchmark points, where the strings persist for much longer, pushing the spectrum down to much lower frequencies.

\begin{figure}
    \centering
    \includegraphics[width=0.8\linewidth]{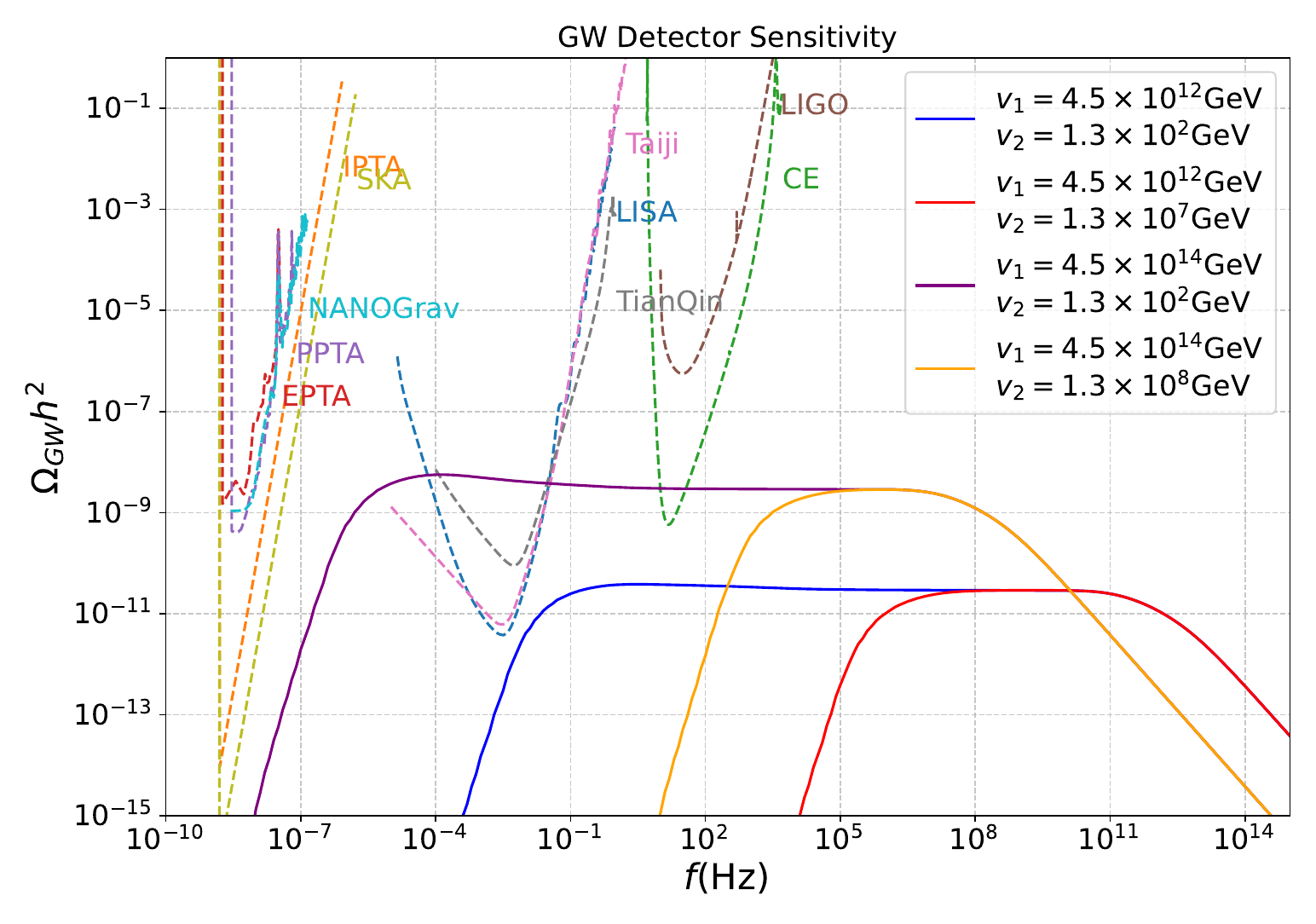}
    \caption{Gravitational wave spectrum of the ``walls eating strings'' scenario in the $v_1 \gg v_2$ limit. In this hierarchical regime the defects can be treated as ideal pure domain walls and pure strings, and the spectra exhibit a cutoff at high frequency. A spectrum sourced by a high-tension string and a low-tension wall is comparatively easier to detect.}
    \label{LargeV}
\end{figure}

In the hierarchical limit $v_1 \gg v_2$, the symmetry breaking reverts to the usual two-step process. Here the ``strings eating walls'' scenario is no longer available: $\kappa_s = \mu^3/\sigma^2 \sim \left( v_1/v_2 \right)^6 \gg 1$, so the domain wall lifetime $t_{\Gamma}$ of Eq.~\eqref{Life_Wall} vastly exceeds the age of the universe, reinstating the usual domain wall problem. Only the ``walls eating strings'' scenario remains viable, and Fig.~\ref{LargeV} shows the resulting gravitational wave spectra for a representative benchmark point, for comparison with the $v_1 \sim v_2$ region that is our primary interest.

\section{Summary and Future Prospect}\label{part 6}

By solving the one-dimensional profiles, relying on a 3:1 U(1) toy model, we dive into the $v_1 \sim v_2$ limit to scrutinize the detailed disappearance processes of the ``$Z_3$ domain wall''. The vacua separated by the meta-stable domain wall deviate from the rigorous $Z_3$ symmetry to a non-negligible scale as the $R_{1 2} = \frac{v_1}{v_2}$ decreases. This deviation can be described by the bias angle $\beta$. Finally the meta-stable domain wall solution converges with another unstable domain wall solution, leaving us no domain wall below some $R_{12}$ scale.

The bias angle $\beta$ should be considered when solving the radius profile of the string-like object in a hybrid wall-string network. These ``strings'' can be classified into ``rims'' and ``conjunctions''. The profiles of the ``rim'' strings, either bounding the domain wall, or nucleated as a hole edge at the domain wall, are calculated.

Since $v_1 \sim v_2$, the universe follow a one-step symmetry breaking. We plot the stochastic gravitational wave spectrum for the ``strings eating walls'', ``walls eating strings'' and ``no wall'' scenarios respectively. However, these preliminary evaluations depend on different assumptions, which are undetermined from our one-dimensional static simulations. Answering these questions requires higher-dimensional dynamic simulations, which has been planned as one of our future research directions. In this paper, the $U(1)$ symmetry and the scalar fields are only toy model elements, and no connections with the SM sector are imposed, which are definitely required by further practical studies. For an example, in this model, the domain wall provides a mutation in the $CP$-phase and a deviation from the thermal equilibrium as a particle passes through it, triggering the possibility to generate the baryon-asymmetry \cite{Brandenberger:1994mq,Sassi:2024cyb,Azzola:2024pzq,Azzola:2026cwa,Vanvlasselaer:2026fay,Bai:2026udq}. We also plan to study some more pragmatic models in such a direction.

\begin{acknowledgements}
    We thank to Ye-Ling Zhou, Zhao-Huan Yu, Zhao zhang for helpful discussions. YLT and JJH
are supported by NSFC under Grant No. 12005312.
BF acknowledges Liaoning Natural Science Foundation No.~2025-BS-0086 and Guangdong Basic and Applied Basic Research Foundation No.~2025A1515011079.
\end{acknowledgements}

\bibliography{reference}

\end{document}